\documentclass[twocolumn,prb,floatfix]{revtex4-1}
\usepackage[dvipdfmx]{graphicx}
\usepackage{amsmath,amssymb}
\usepackage{bm}
\usepackage{makeidx}
\usepackage{float}
\makeindex

\begin{document}
\author{Karin Matsumoto, Daisuke Ogura, and Kazuhiko Kuroki}
\affiliation{Department of Physics, 
Osaka University, 1-1 Machikaneyama, 
Toyonaka, Osaka 560-0043, Japan}
\title{Wide applicability of high $T_c$ pairing originating from coexisting wide and incipient narrow bands in quasi one dimension}
\begin{abstract}
We study superconductivity in the Hubbard model on various quasi-one-dimensional lattices with coexisting wide and narrow bands originating from multiple sites within a unit cell, where each site corresponds to a single orbital. The systems studied are the two-leg and three-leg ladders, the diamond chain, and the criss-cross ladder. These one-dimensional lattices are weakly coupled to form two-dimensional (quasi-one-dimensional) ones, and the fluctuation exchange approximation is adopted to study spin-fluctuation-mediated superconductivity. When one of the bands is perfectly flat, and the Fermi level, intersecting the wide band, is placed in the vicinity of, but not within, the flat band, superconductivity arising from the interband scattering processes is found to be strongly enhanced owing to the combination of the light electron mass of the wide band and the strong pairing interaction due to the large density of states of the flat band. Even when the narrow band has finite band width, the pairing mechanism still works since the edge of the narrow band, due to its large density of states, plays the role of the flat band. The results indicate the wide applicability of the high $T_c$ pairing mechanism due to coexisting wide and ``incipient'' narrow bands in quasi-one-dimensional systems.
\end{abstract}
\maketitle

\section{Introduction}
More than thirty years have passed since the discovery of the high $T_c$ cuprate superconductors, but the cuprates remain to be the record holder of the superconducting transition temperature $T_c$ at ambient pressure. 
A difficulty in realizing high $T_c$ superconductivity lies in the incompatibility between a strong pairing interaction and a light renormalized electron mass: 
a strong electron correlation can mediate strong pairing interactions, but also makes the electron mass heavy due to the strong quasiparticle renormalization. The former is of course favorable for superconductivity, but the latter suppresses $T_c$.   For instance, it is known that in the single-band Hubbard model on a square lattice, a model for the high-$T_c$ cuprates, the pairing interaction is the largest at half-filling, but there the system is a Mott insulator \cite{White_QMC}. 

About a decade ago, one of the present authors proposed a way to overcome this problem\cite{Kuroki2005_wide-narrow}. The proposal requires a system where narrow and wide bands coexist, not due to the presence of multiple orbitals (like $f$ and $s$ orbitals\cite{Suhl,Kondo}), but due to the presence of multiple sites within a unit cell. The case where the narrow band intersects the wide band is considered,  and the Fermi level is set close to, but not within, the narrow band. 
Then the electrons in the wide band, which are not so strongly renormalized and have light effective mass, can form Cooper pairs with a strong pairing interaction mediated by the large number of interband scattering channels originating from the large density of states of the narrow band (see upper panel of Fig.\ref{fig1}). An important point is that the large density of states of the narrow band does not cause strong  renormalization since it does not intersect the Fermi level. Here it is worth mentioning a recent trend of the studies of the iron-based superconductors, in which bands that do not intersect, but lie close to, the Fermi level has been referred to as the ``incipient band'', and the possibility of enhanced superconductivity due the interband scattering processes involving such bands has been investigated intensively\cite{Hirschfeld2011_incipient, Miao2015_ironbased111FS, Wang2011_122FRG, Bang2014_122shadowgap, Chen2015_incipient, Bang2016_dynamicaltuning}. These studies have been  motivated by the experimental observation of the hole band sinking below the Fermi level in some of the iron-based superconductors\cite{Guo2010_ironbased122, Qian2011_ironbased122FS, Wang2012_ironbased11-STO, Tan2013_ironbased11-STO, Miao2015_ironbased111FS, Niu2015_ironbased11, Mishra2016_bilayer}.  The narrow band considered in Ref.\onlinecite{Kuroki2005_wide-narrow} lying in the vicinity of the Fermi level is indeed the incipient band in today's terminology.

Coming back to Ref.\onlinecite{Kuroki2005_wide-narrow}, the two-leg Hubbard ladder (Fig.\ref{fig1}(a)) was considered as an actual system in which the above mentioned high $T_c$ mechanism works. The Hubbard ladder had already been studied intensively after the proposal by Dagotto and Rice\cite{Dagotto1992_ladder,Rice1993_ladder,Dagotto-Rice1996_ladder}, and superconductivity with a $T_c$ of above 10K was indeed observed in (Sr,Ca)$_{14}$Cu$_{24}$O$_{41}$\cite{Nagata1998_14-24-41}. The proposal in Ref.\onlinecite{Kuroki2005_wide-narrow} was that 
a much higher $T_c$ may take place in two-leg cuprate ladder compounds, if $\sim 30$ \% of electron doping can be achieved.
Namely, in the two-leg ladder lattice with only the nearest-neighbor hoppings $t$, there are the bonding and antibonding bands with the same band widths separated by $2t$. When the second-neighbor (diagonal) hoppings $t'$ are introduced, one of the bands become narrower, while the other becomes wider. About 30 \% of electron doping will place the Fermi level just above the narrow band, making the narrow band ``incipient''.

More recently, another model on the so-called ``diamond chain'', shown in Fig.\ref{fig1}(c), has been studied as a system with coexisting narrow (flat) and wide bands\cite{Kobayashi2016_flatband}. There, applying exact diagonalization and density matrix renormalization group to a finite cluster, it has been shown that superconductivity is strongly enhanced when the Fermi level is positioned close to, but not within, the flat band. This study, combined with Ref.\onlinecite{Kuroki2005_wide-narrow}, suggests generality of enhanced superconductivity in systems with coexisting wide and incipient narrow bands.

In fact, we can expect that the mechanism may work in yet another system, namely, the three-leg Hubbard ladder, shown in Fig.\ref{fig1}(b). Theoretically it has been known that the three-leg Hubbard ladder also exhibits superconductivity\cite{Arrigoni1996_3-leg,Schulz1996_ladder,Kimura1996_3leg-ladder,Lin-Balents-Fisher1997_n-leg}.  In those studies, the effect of the diagonal hoppings was not explicitly studied, but the introduction of such hoppings, as in the two-leg case, makes the width of the three bands vary from the widest to the narrowest, so that an enhancement of superconductivity due to the coexistence of wide and incipient narrow bands may also be possible in the three-leg ladder model.

Given this background, here we study the Hubbard models on the above mentioned lattices with coexisting wide and narrow bands, and investigate how superconductivity is affected depending on the relation between the Fermi level and the narrow band energy. These lattices themselves are purely one-dimensional, but here we consider cases where these one-dimensional lattices are weakly coupled to form two-dimensional ones, and apply the fluctuation exchange approximation.  In the cases when the narrow band is perfectly flat, it is found in all models that superconductivity is enhanced when the Fermi level is placed close to, but not within, the flat band. When the narrow band possesses finite width, the band edge plays the role of the flat band since the density of states at the band edge is large in quasi-one-dimensional systems. The present study suggests the wide applicability of the high $T_c$ mechanism owing to coexisting wide and incipient narrow bands.

\section{The models and methods}

\begin{figure}
\includegraphics[width=7.5cm]{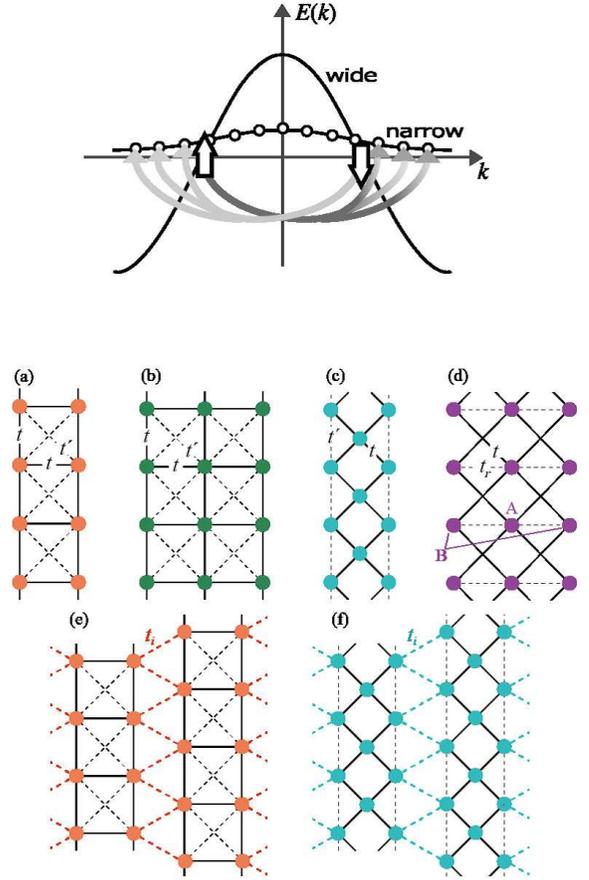}
\caption{Upper panel : schematic image of coexisting wide and incipient narrow bands. Lower panels : models in real space. (a) two-leg ladder, (b) three-leg ladder, (c) diamond chain, (d) criss-cross ladder, (e) and (f) show the coupling between the ladders and the diamond chains, respectively.} 
\label{fig1}
\end{figure}

\begin{figure}
\begin{minipage}{0.48\hsize}
\centering
\includegraphics[width=4.2cm] {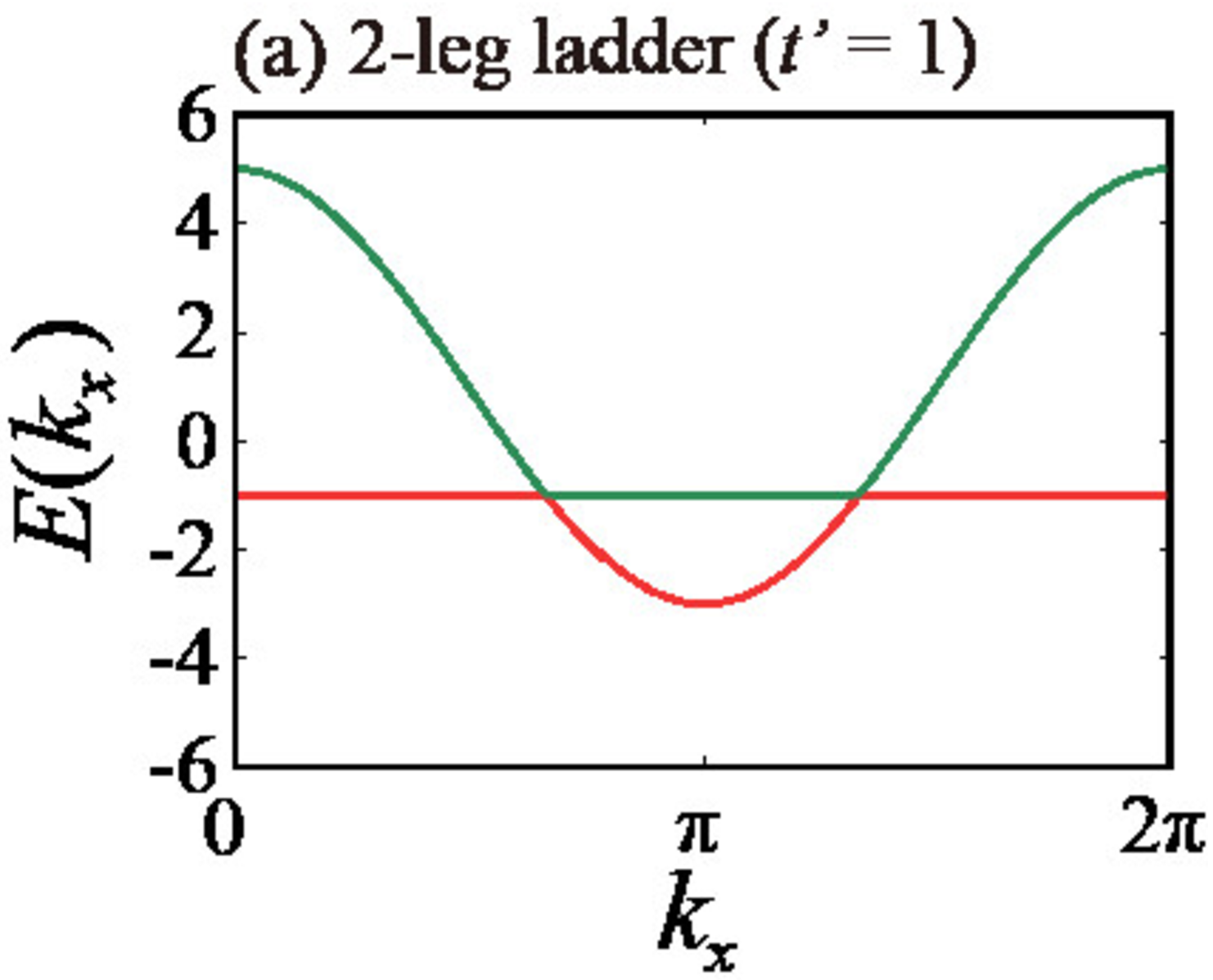}
\end{minipage}  
\begin{minipage}{0.48\hsize}
\includegraphics[width=4.2cm] {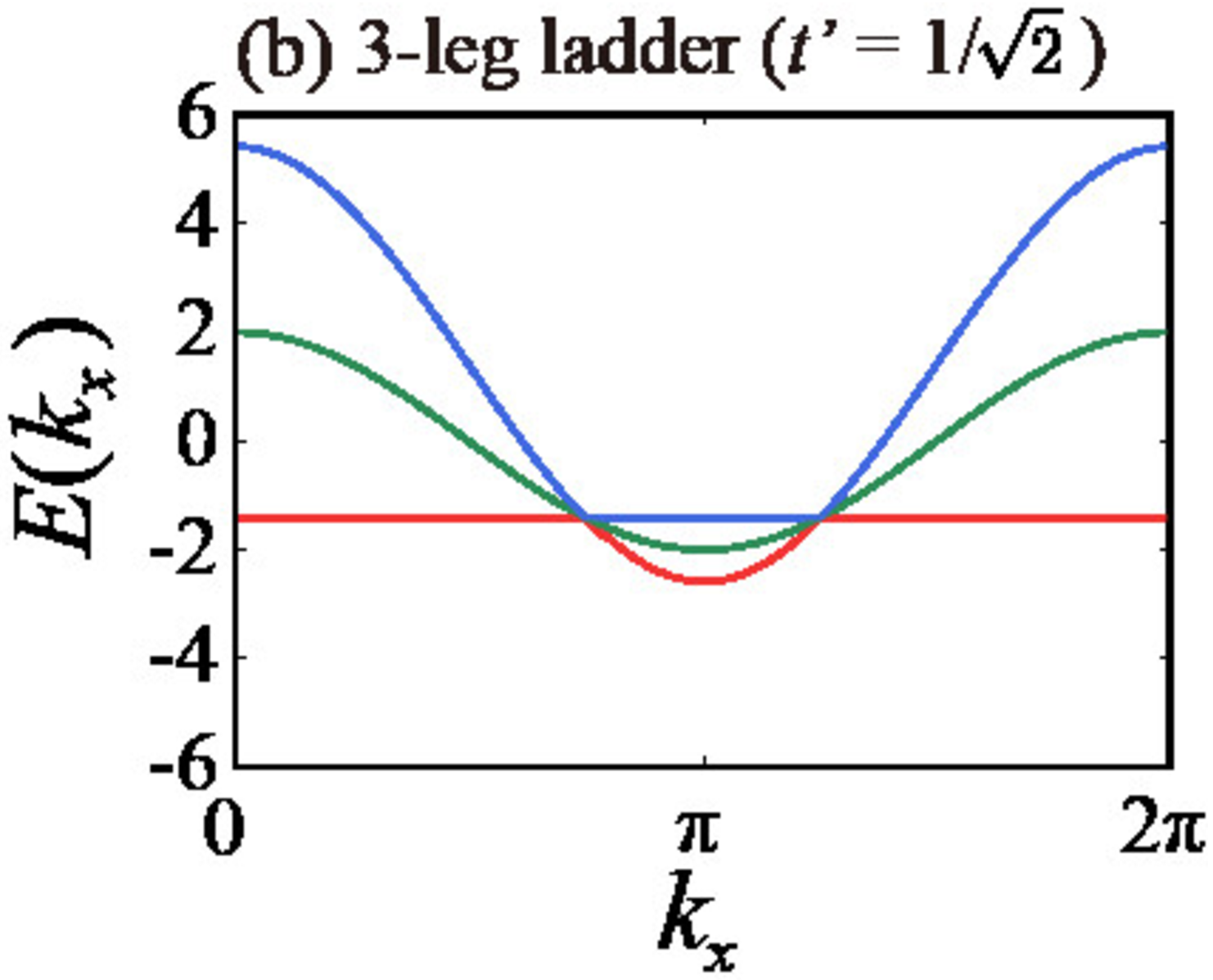}
\end{minipage}
\\
\begin{minipage}{0.48\hsize}
\includegraphics[width=4.2cm] {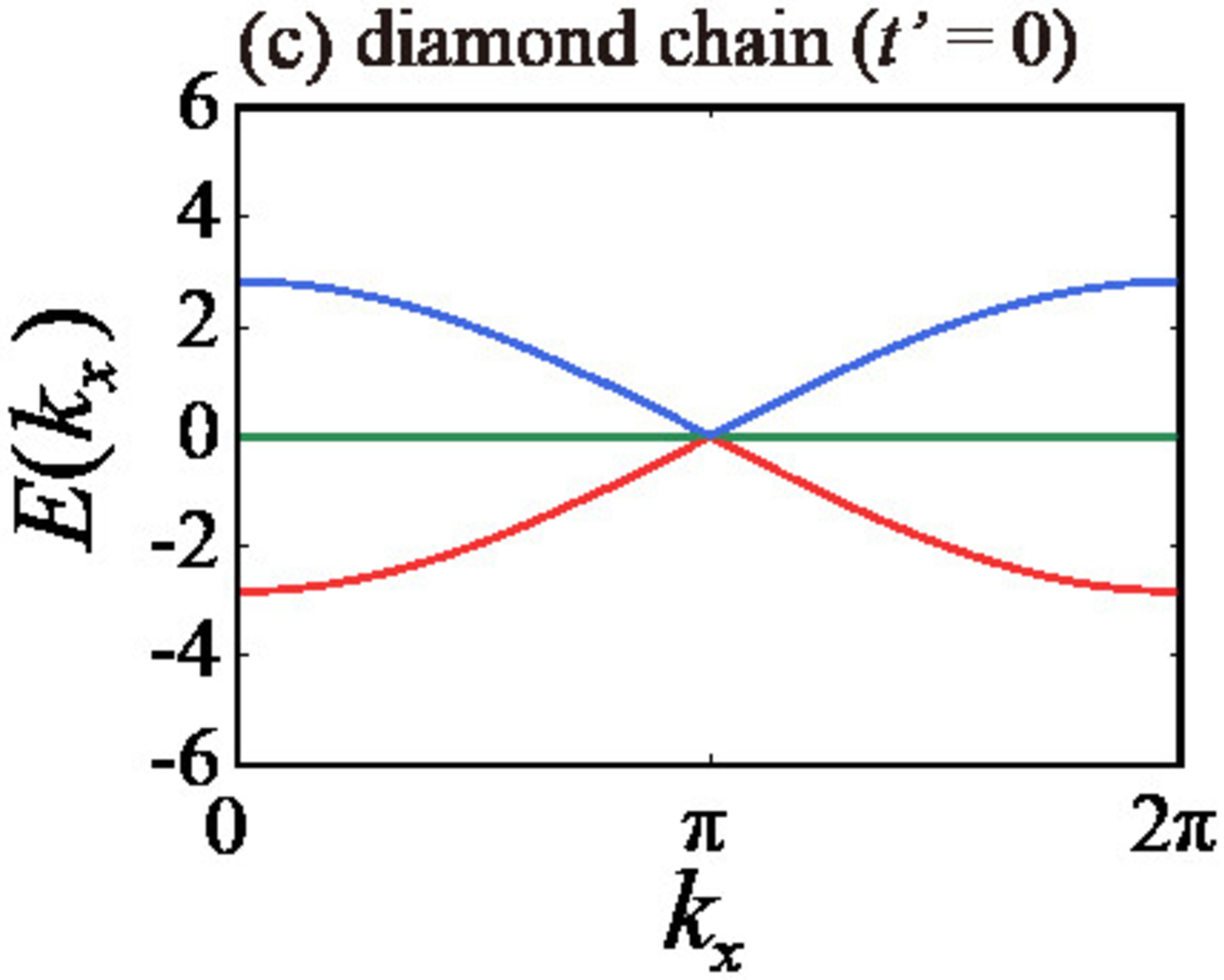}
\end{minipage}   
\begin{minipage}{0.48\hsize}
\includegraphics[width=4.2cm] {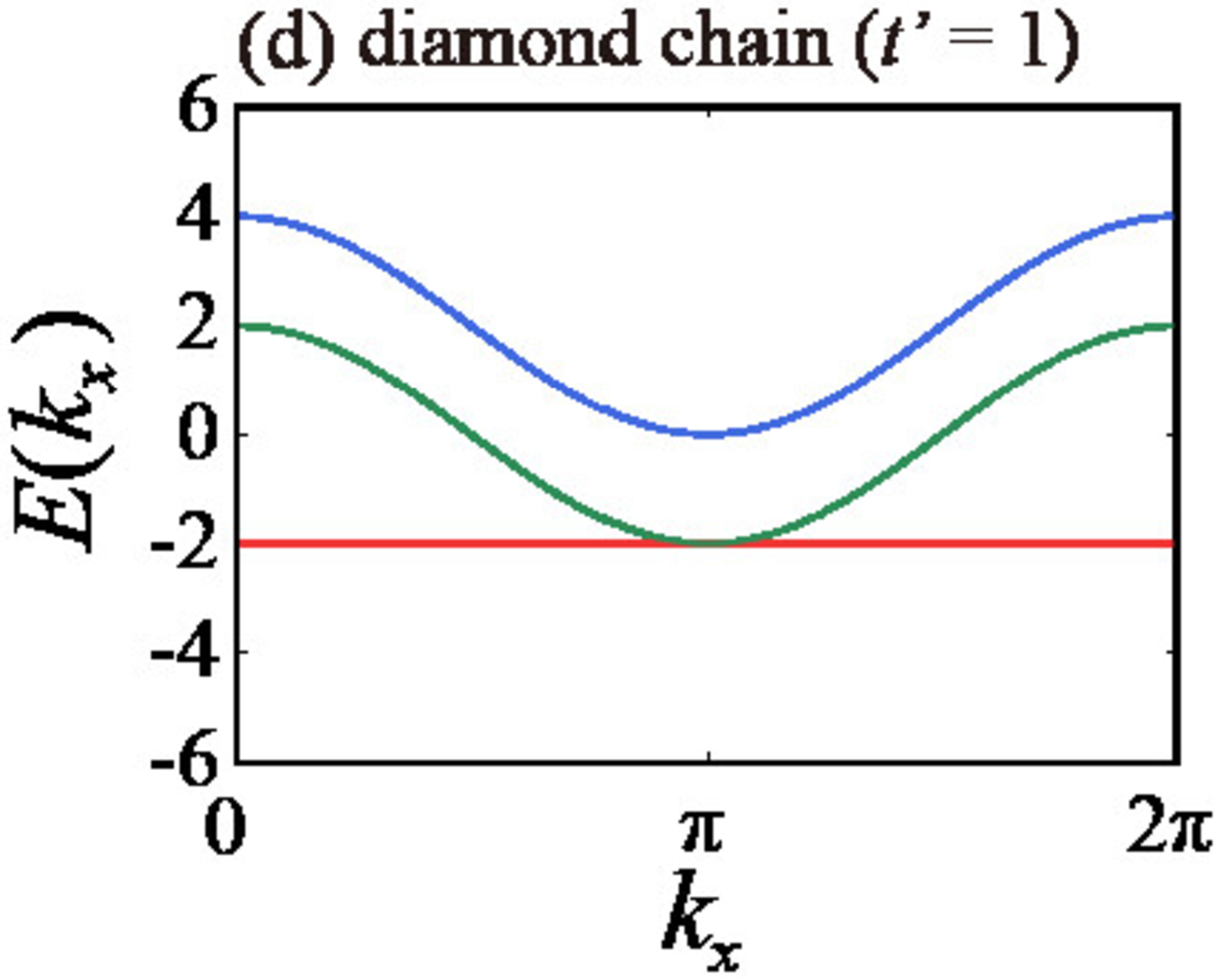}
\end{minipage}
\begin{minipage}{0.48\hsize}
\centering
\includegraphics[width=4.2cm] {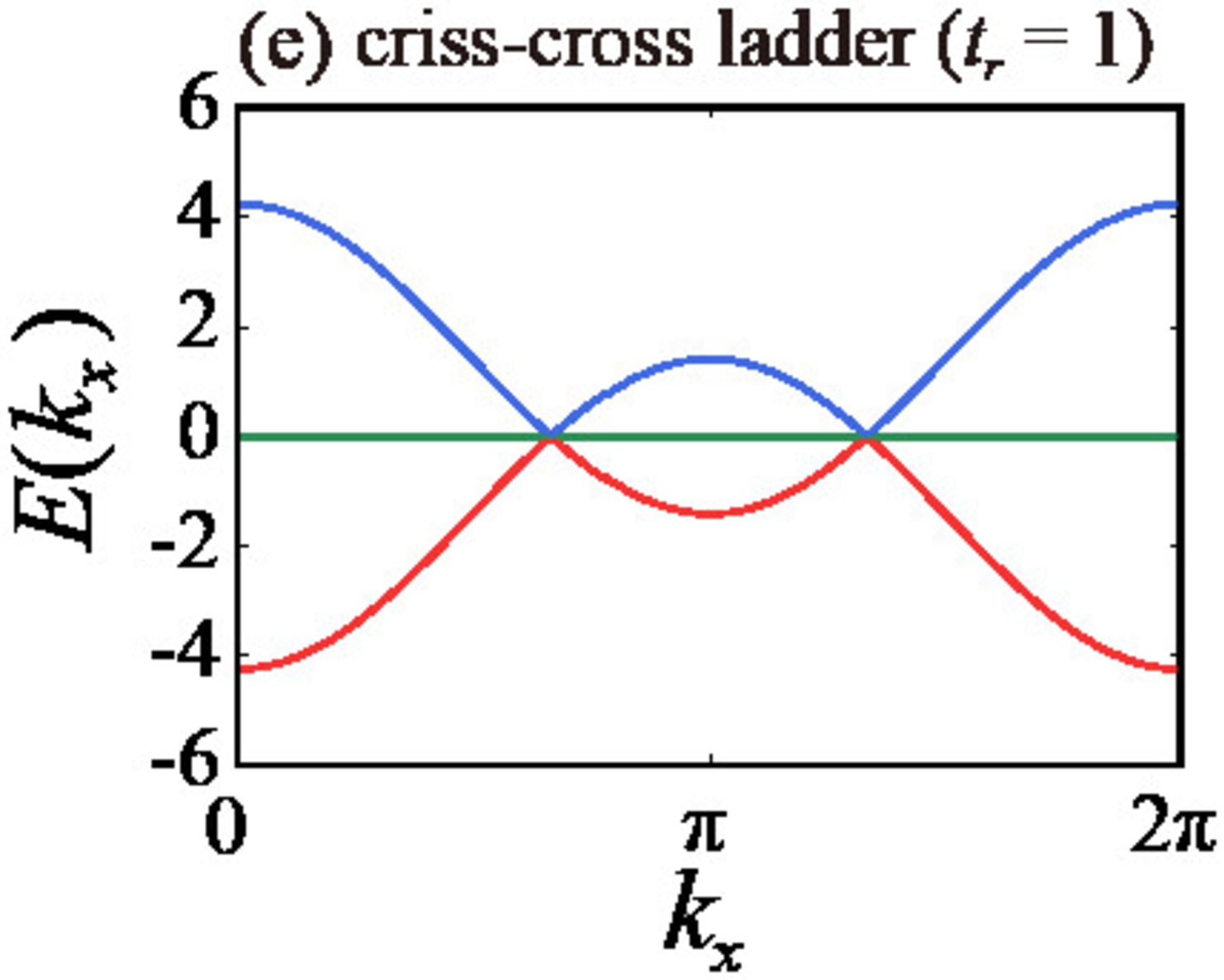}
\end{minipage}
\begin{minipage}{0.48\hsize}
\centering
\includegraphics[width=4.2cm] {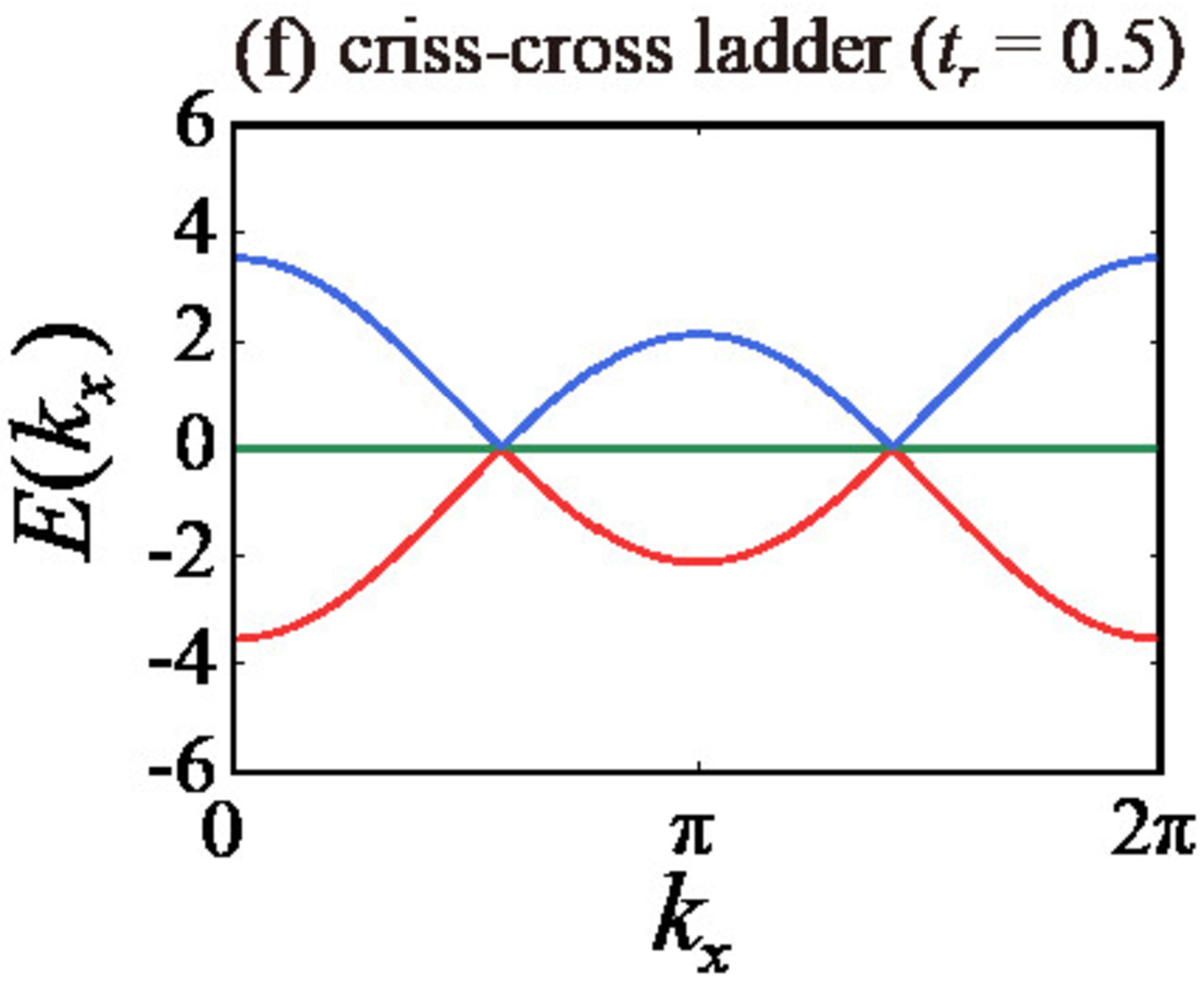}
\end{minipage}
\begin{minipage}{0.48\hsize}
\centering
\includegraphics[width=4.2cm] {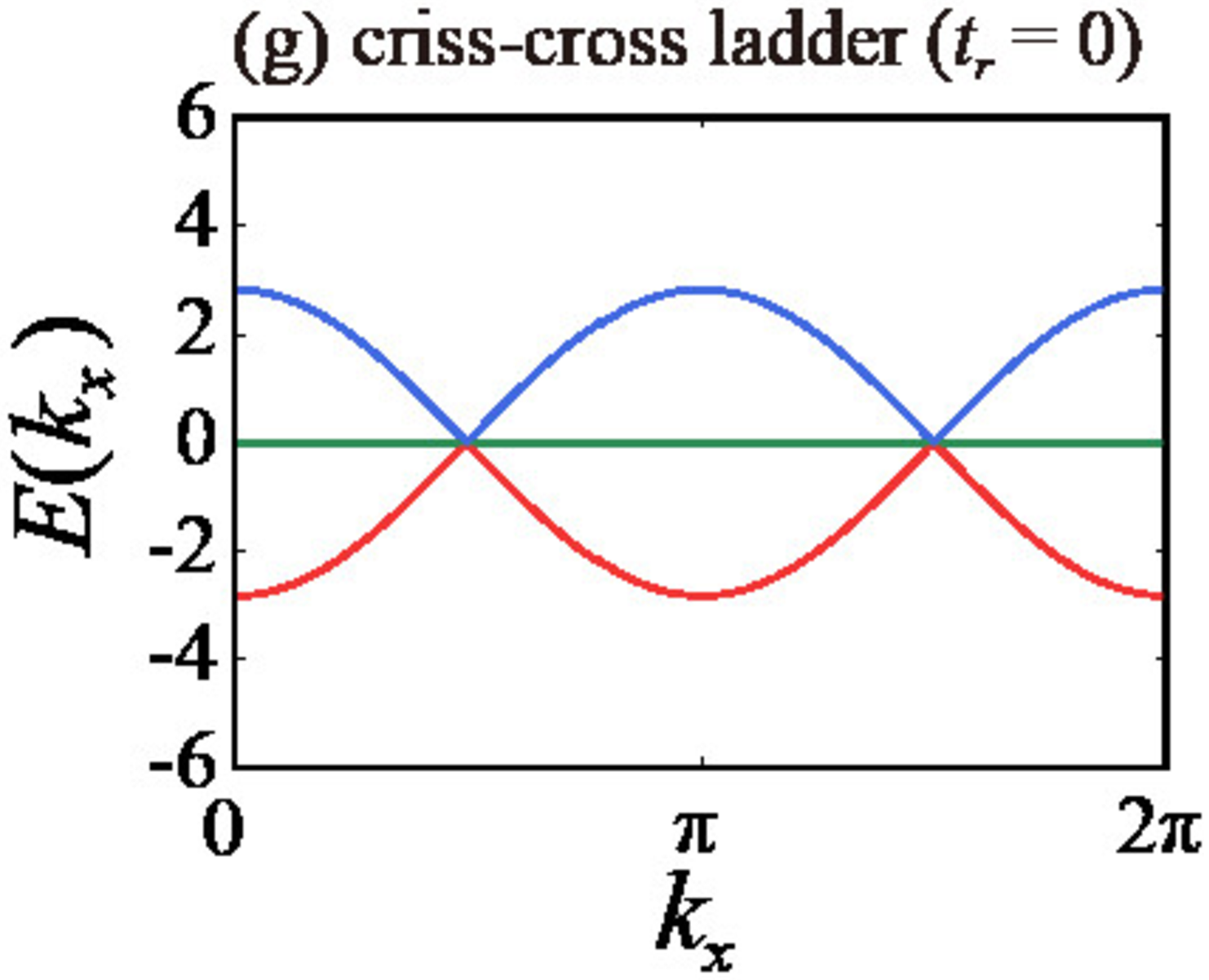}
\end{minipage}
\caption{The bare energy bands of the tightbinding models having flat bands. (a) two-leg ladder with $t'=1$, (b) three-leg ladder with $t'=1/\sqrt{2}$, diamond chain with (c) $t'=0$ and (d) $t'=1$, criss-cross ladder with (e) $t_r=1$, (f) $t_r=0.5$, and (g) $t_r=0$.}
\label{fig2}
\end{figure}

We start with describing the tightbinding lattices considered in the present study. All the models are single orbital models, namely, there is only one orbital per site. The nearest neighbor hoppings are $t$ and taken as the unit of the energy $(t=1)$ (except for the criss-cross ladder described below), and the second nearest neighbor hoppings are $t'$. $x$-axis is taken as the direction of the chains. $c_{k_x,\sigma}$ etc. are annihilation operators of an electron with momentum $k_x$ and spin $\sigma$, and $c$, $d$, etc. denote the different sites within a unit cell.  In the actual calculation, all the models are weakly coupled to form two-dimensional lattices, but we first describe their purely one-dimensional forms.

In the two-leg ladder, shown in Fig.\ref{fig1}(a), the Hamiltonian takes the form 
\footnotesize
\begin{align}
{\cal H}_{k}^{{\footnotesize 2\mathchar `- leg}} &= \sum_{k_x,\sigma} 
\begin{pmatrix}
c_{k_x,\sigma}^\dag & d_{k_x,\sigma}^\dag 
\end{pmatrix} \nonumber \\
&\begin{pmatrix}
2t \mathrm{cos}k_x  &t +2t' \mathrm{cos}k_x \\
t +2t' \mathrm{cos}k_x  & 2t \mathrm{cos}k_x  
\end{pmatrix} 
\begin{pmatrix}
c_{k_x,\sigma} \\
d_{k_x,\sigma}  
\end{pmatrix}. 
\label{2lad-ham}
\end{align}
\normalsize
The diagonalized energy bands take the form,
\footnotesize
\begin{equation}
E_{\pm}(k_x)= (t \mp t^\prime) \mathrm{cos} k_x \mp t \label{2lad-energy}.
\end{equation}
\normalsize
When $t'=1$, one of the bands become perfectly flat as shown in Fig.\ref{fig2}(a).

In the three-leg ladder, shown in Fig.\ref{fig1}(b), the Hamiltonian is given as 
\footnotesize
\begin{align}
{\cal H}_{k}^{{\footnotesize 3\mathchar `- leg}} &= \sum_{k_x,\sigma} 
\begin{pmatrix}
c_{k_x,\sigma}^\dag & d_{k_x,\sigma}^\dag & e_{k_x,\sigma}^\dag
\end{pmatrix}  \nonumber \\
&\begin{pmatrix}
2t \mathrm{cos}k_x  & t +2t^\prime \mathrm{cos}k_x  &0 \\
t +2t^\prime \mathrm{cos}k_x  & 2t \mathrm{cos}k_x  &t +2t^\prime \mathrm{cos}k_x\\
0 & t +2t^\prime \mathrm{cos}k_x & 2t \mathrm{cos}k_x 
\end{pmatrix}  
\begin{pmatrix}
c_{k_x,\sigma} \\
d_{k_x,\sigma} \\
e_{k_x,\sigma}  \label{3lad-ham}
\end{pmatrix},
\end{align}
\normalsize
and the diagonalized energy bands are 
\footnotesize
\begin{align}
E_{3 \mathchar `- leg,1}(k_x) &= 2t \mathrm{cos} k_x , \\ \nonumber
E_{3 \mathchar `- leg ,\pm}(k_x) &= 2(t \pm  \sqrt{2} t^\prime) \mathrm{cos} k_x  \pm \sqrt{2}t. 
\end{align}
\normalsize
When $ t^\prime = 1/\sqrt{2} $, one of the bands become perfectly flat as shown in Fig.\ref{fig2}(b).

The Hamiltonian of the diamond chain is 
\footnotesize
\begin{align}
{\cal H}_{k}^{{\footnotesize dia}} &= \sum_{k_x,\sigma} 
\begin{pmatrix}
c_{k_x,\sigma}^\dag & d_{k_x,\sigma}^\dag & e_{k_x,\sigma}^\dag
\end{pmatrix} \nonumber \\
&\begin{pmatrix}
2t' \mathrm{cos}k_x  &t(1 + e^{ik_x})  &0   \\
t(1 + e^{-ik_x}) &0 &t(1 + e^{-ik_x})     \\
0 &t(1 + e^{ik_x}) &2t' \mathrm{cos}k_x   
\end{pmatrix}
\begin{pmatrix}
c_{k_x,\sigma} \\
d_{k_x,\sigma} \\
e_{k_x,\sigma} 
\end{pmatrix}, \label{dia-ham}
\end{align}
\normalsize

\noindent
and the energy bands are given as 
\footnotesize
\begin{align}
E_{dia,1}(k_x) &= 2 t^\prime \mathrm{cos} k_x , \\ \nonumber
E_{dia,2,3}(k_x) &=  t^\prime \cos{k_x} \pm \sqrt{t^{\prime 2} \cos^2{k_x}
+4 t^2 (1 + \cos{k_x} )}. 
\end{align}
\normalsize
When $t^\prime = 0$ or $t^\prime = 1$, one of the bands become perfectly flat as shown in Fig.\ref{fig2}(c) and (d), respectively.

We can further consider a lattice that always has a flat band, and provides a link between the three-leg ladder and the diamond chain. Namely, if we remove the hoppings of the three-leg ladder in the leg direction  and leave only the hopping in the rung direction (denoted as $t_r$ here) and the the diagonal hoppings (denoted  as $t$ here),  we end up with the lattice shown in Fig.\ref{fig1}(d). We will call this lattice the ``criss-cross ladder''. If we assign A and B sublattices in this lattice as in the figure, the number of A and B sites within a unit cell differs by one, and all the hoppings connect A and B sublattice sites, so that there is always a flat band regardless of the values of the hoppings according to Lieb's theorem\cite{Lieb}. When the hopping in the rung direction is taken as $t_r=0$, this lattice is nothing but the diamond chain (two decoupled diamond chains).
In fact, the Hamiltonian of this lattice is 

\footnotesize
\begin{align}
{\cal H}_{k}^{{\footnotesize c \mathchar `- c}} &= \sum_{k_x,\sigma} 
\begin{pmatrix}
c_{k_x,\sigma}^\dag & d_{k_x,\sigma}^\dag & e_{k_x,\sigma}^\dag
\end{pmatrix}  \nonumber \\
&\begin{pmatrix}
0 & t_r +2t \mathrm{cos}k_x  &0 \\
t_r +2t \mathrm{cos}k_x  & 0  &t_r +2t \mathrm{cos}k_x\\
0 & t_r +2t \mathrm{cos}k_x & 0
\end{pmatrix}  
\begin{pmatrix}
c_{k_x,\sigma} \\
d_{k_x,\sigma} \\
e_{k_x,\sigma}  \label{crisscross-ham}
\end{pmatrix},
\end{align}
\normalsize
and the energy bands are 

\footnotesize
\begin{align}
E_{c \mathchar `- c,1}(k_x) &=0, \\ \nonumber
E_{c \mathchar `- c,\pm}(k_x) &= \pm 2\sqrt{2} t \mathrm{cos} k_x  \pm \sqrt{2}t_r. 
\end{align}
\normalsize
These equations confirm the above mentioned properties.
In order to strengthen the generality of the mechanism, we also study this lattice in the present study, fixing $t=1$ and varying $t_r$.

As mentioned, all the lattices are weakly connected in the $y$ direction to form two dimensional lattices. The coupling  $t_i$ between the one-dimensional lattices are introduced in the form shown in Fig.\ref{fig1}(e)fe) as in the actual two-leg and three-leg cuprate ladder compounds\cite{Kontani_trellis}. In the actual two-leg compound $\mathrm{SrCu_2O_3}$,  $t_i/t$ varies from 0.03 to 0.07 depending on how to evaluate\cite{Muller_SrCu2O3}, so we took $t_i/t =0.1$ in the present study. In all cases, the band filling $n$ is defined as the average number of electrons per unit cell (maximum is $n=4$ for the two-leg ladder, and $n=6$ for the three-leg ladder, the diamond chain, and the criss-cross ladder).

On top of these tightbinding models, we consider the on-site repulsive Hubbard interaction $U$. Throughout the paper, $U=6$ is adopted, which is a typical value (in units of $t$) for the cuprates and related transition metal oxides\cite{Vaugier_2012_cRPA_3d-4d,Mravlje_2011_cRPA,Jang_2016_cRPA}. We apply the fluctuation  exchange (FLEX) approximation\cite{Bickers1991} to obtain the renormalized Green's function. Namely, bubble and ladder type diagrams are collected to obtain the spin and charge susceptibilities, which enter the effective interaction that is necessary to obtain the self energy. The Dyson's equation is solved using the self energy, which gives the renewed Green's function, and the self energy is recalculated. This iteration process is repeated till convergence is attained. The Green's function is first obtained in the site representation, namely, in the form of $G_{\alpha\beta}$, where $\alpha$, $\beta$ denotes the sites within a unit cell. Then it is transformed into the band representation by a unitary transformation.
To study superconductivity mediated by the spin fluctuation, the converged Green's function and the susceptibilities are plugged into the linearized Eliashberg equation, 
whose eigenvalue $\lambda$ reaches unity at the superconducting transition temperature $T = T_c$, so that it can be considered as a measure of $T_c$  when calculated at a fixed temperature. Unless noted otherwise, we calculate the eigenvalue at $T=0.05$, where $32\times 32$ $k$-point mesh and 1024 Matsubara frequencies are taken. Larger number of Matsubara frequencies are taken for lower temperature, e.g., for $T=0.025$, 2048.

\section{RESULTS}
\subsection{Cases with a perfectly flat band}
\label{flatcase}
\begin{figure}
\includegraphics[width=7.5cm]{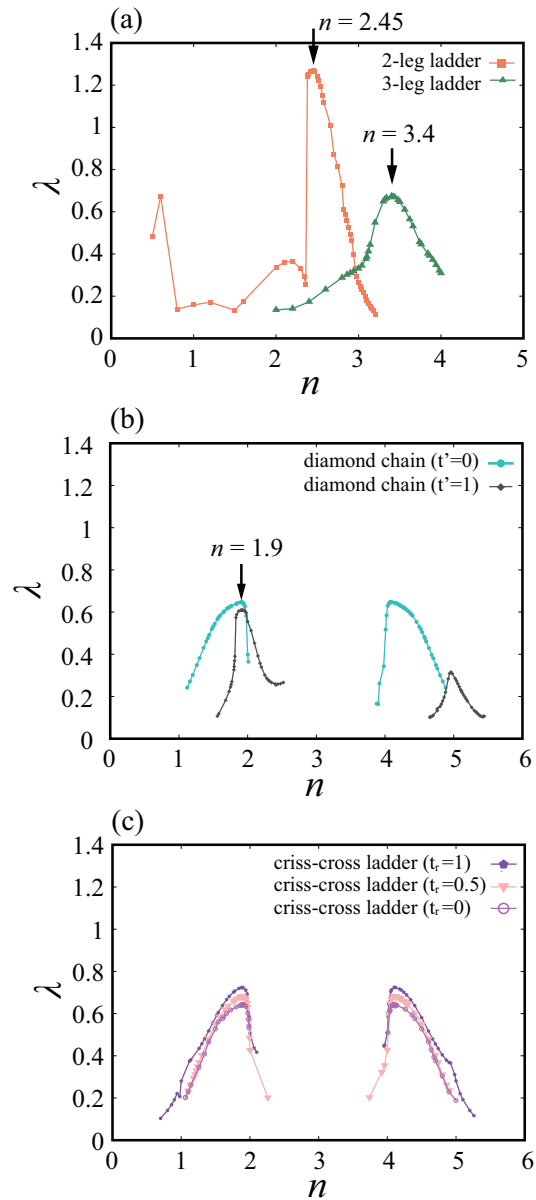}
\caption{$\lambda$ against the band filling for (a) the two-leg and three-leg ladders, (b) the diamond chain with $t'=0$ or $t'=1$, (c) the criss-cross ladder with $t_r=1, 0.5, 0$. The linearized Eliashberg equation could not be solved at the band fillings where $\lambda$ is not plotted; there $\lambda$ is expected to be very small. }
\label{fig3}
\end{figure}

\begin{figure}
\includegraphics[width=8.5cm]{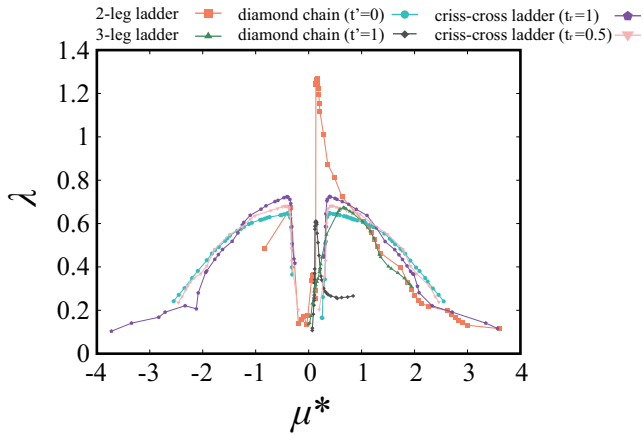}
\centering
\caption{$\lambda$ against $\mu^*$ (see text for definition) for all flat-band cases}
\label{fig4}
\end{figure}

In this section,  we consider cases where the bare ($U=0$) narrow band is perfectly flat in the chain ($k_x$) direction (i.e., consider the band structures shown in Fig.\ref{fig2}). We first show the band filling dependence of the eigenvalue of the Eliashberg equation. The results for the two-leg and three-leg ladders, those for the diamond chain, and those for the criss-cross ladder are given in Fig.\ref{fig3}(a),(b), and (c), respectively. It can be seen that in all cases, superconductivity is sharply optimized at a certain band filling. In order to elucidate the role of these band fillings, we replot in Fig.\ref{fig4} the eigenvalue against chemical potential (Fermi level). Here, we use the chemical potential $\mu$ for the non-interacting case ($U=0$), and define a chemical potential measured from the energy of the flat band $E_{\rm flat}$ as 
\begin{equation}
\footnotesize
\mu^*=(\mu-E_{\rm flat})\times\frac{W_{\rm 2leg}}{W}
\end{equation}
Here, $W$ is the total band width in the non-interacting case, and $W_{\rm 2leg}$ in particular is that for the two-leg ladder. The factor $\frac{W_{\rm 2leg}}{W}$ is to take into account the difference in the total band width among different systems. In all of the systems, the eigenvalue is maximized when the chemical potential lies in the vicinity of the flat band.  This result shows that the high $T_c$ pairing  mechanism in coexisting wide and incipient narrow bands works rather generally among various quasi-one-dimensional systems. It is also important to notice that superconductivity is strongly suppressed when the chemical potential is too close to the flat band. In other words, superconductivity occurs rather abruptly against the variation of the chemical potential or the electron density. 

The optimized superconductivity has extremely high $T_c$ especially for the two-leg ladder case, considering the fact that $\lambda$ already exceeds unity for the temperature adopted here, 0.05, in units of the nearest neighbor hopping ($\sim$300K assuming a typical value of nearest neighbor hopping $\simeq$ 0.5eV). We note here that the FLEX approximation gives a maximum $T_c$ of somewhat lower than 100K for the Hubbard model with a realistic band structure of a high $T_c$ cuprate HgBa$_2$CuO$_4$\cite{Sakakibara2010}. This estimation is close to, or somewhat underestimates the actually observed $T_c$.

In the case of diamond chain with $t'=0$ and the criss-cross ladder, the band structure is electron-hole symmetric, and hence the eigenvalue takes the same values between cases when the chemical potential lies below and above the flat band. In the two-leg ladder, the eigenvalue peak is larger when the chemical potential is above the flat band than when it is below. This is because the band filling is closer to half-filling ($n=2$ in the two-leg ladder) in the former, so that the electron correlation effect is stronger. In the three-leg ladder case also, there can be enhanced superconductivity when the chemical potential lies just below the flat band. However, the band filling in that case is very small, and the linearized Eliashberg equation could not be solved appropriately in that regime.

If we look more closely, there are some other differences among the systems. As mentioned above, the eigenvalue of the two-leg ladder is by far the largest. The reason for this is probably because both of the bands contribute to superconductivity, while in other cases, only one of the dispersive bands is relevant. For example, it is known that the middle band in Fig.\ref{fig2}(b) does not contribute to superconductivity in the three-leg ladder\cite{Kimura1996_3leg-ladder,Schulz1996_ladder,Lin-Balents-Fisher1997_n-leg}. In the diamond chain, one of the dispersive bands does not intersect the Fermi level, and hence should not contribute to  superconductivity. 

Another difference is that the eigenvalue of the two-leg ladder is peaked when the chemical potential lies in the very vicinity of the flat band. By contrast, In the three-leg ladder the eigenvalue takes its maximum at a chemical potential value somewhat away from the flat band position.  To understand the difference among the systems, we plot,  against $k_x$,  the absolute value of the renormalized Green's function in the band representation at the lowest Matsubara frequency. Multiple values are plotted at each $k_x$, which is due to the small but finite dispersion in the $k_y$ direction originating from the interladder(or chain) hopping $t_i$. If a renormalized band intersects the chemical potential at a certain wave vector, the Green's function will be peaked at that wave vector.  Hereafter we will denote the wave vector at which the renormalized (bare) dispersive  band intersects the chemical potential as $k_F^{\mathrm{eff}}$ ($k_F^{(0)}$).

\begin{figure}
\centering
\textbf{(a) $n = 2.7$}
\\
\begin{minipage}{1.0\hsize}
\includegraphics[width=8.5cm]{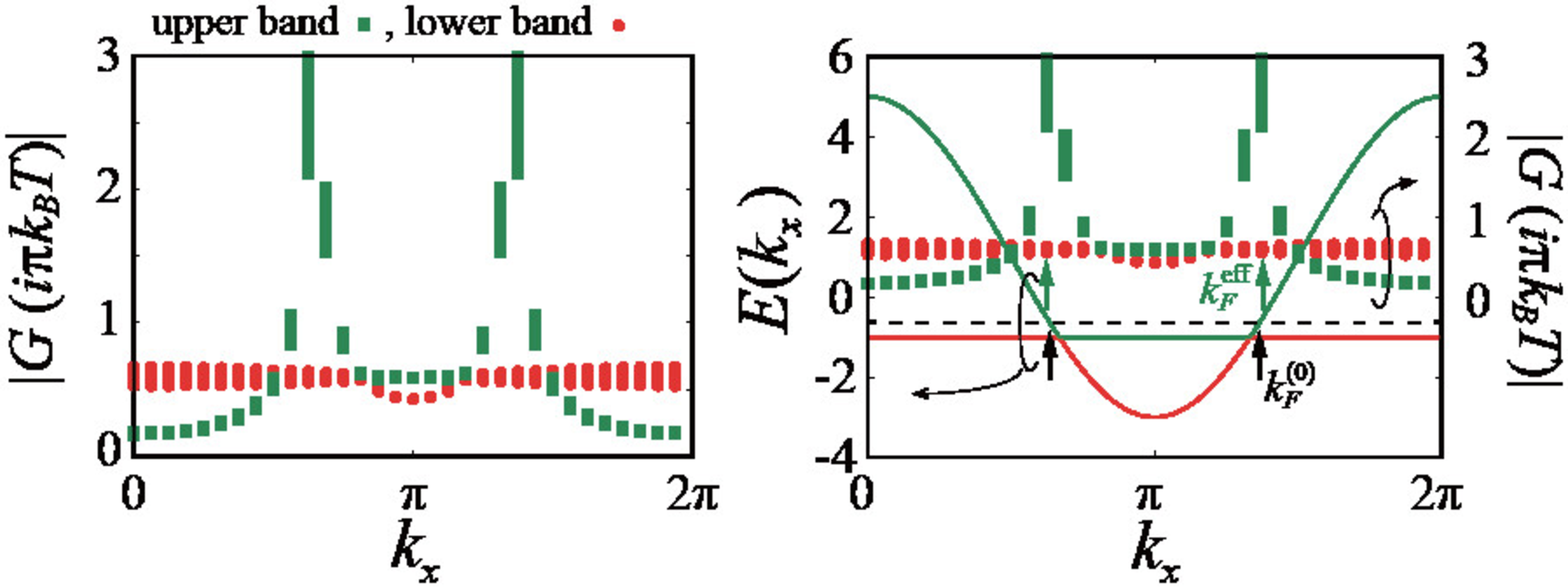}
\end{minipage}
\\
\centering
\textbf{(b) $n = 2.45$}
\\
\begin{minipage}{1.0\hsize}
\includegraphics[width=8.5cm] {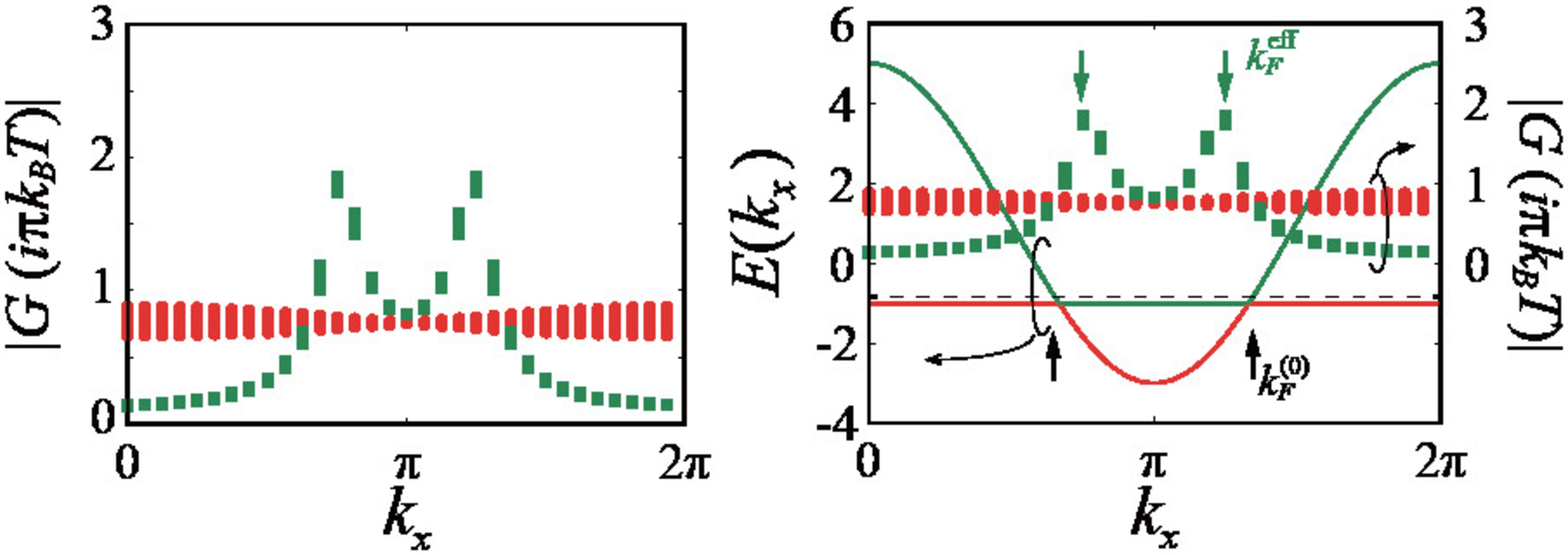}
\end{minipage}
\\
\centering
\textbf{(c) $n = 2.2$}
\\
\begin{minipage}{1.0\hsize}
\includegraphics[width=8.5cm] {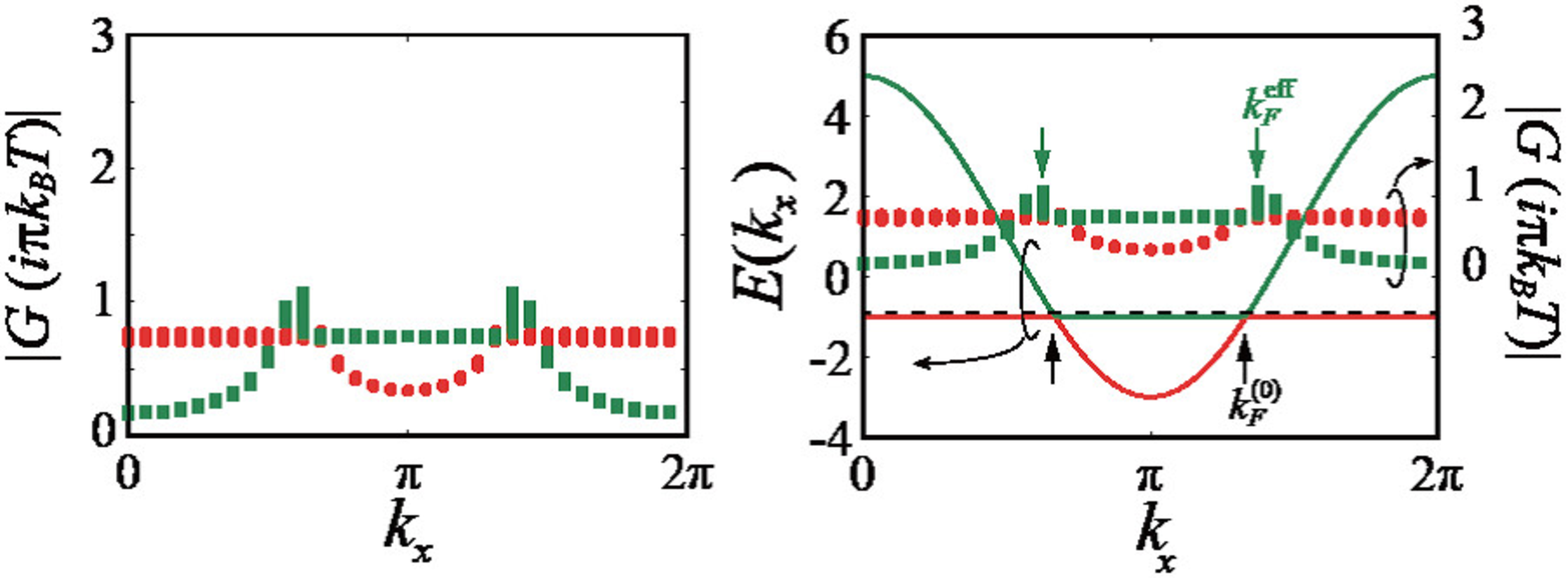}
\end{minipage}
\caption{Green's function for the two-leg ladder plotted against $k_x$. (a) $n=2.7$, (b) $n=2.45$, (c) $n=2.2$. Throughout the paper, the bare band structure is superimposed onto the Green's function by the solid lines on the right panels.  The Green's function of a certain portion of the band is indicated by those with the same color as the band. The dashed line indicates the position of the bare chemical potential. See the text for the definition of $k_F^{(0)}$ and $k_F^{\rm eff}$.}
\label{fig5}
\end{figure}

We start with the two-leg ladder. In Fig.\ref{fig5}, we notice from the small $k_x$ dependence of the Green's function of the flat band that this band remains to be flat even when the interaction is turned on. When $n=2.7$, for which $\lambda$ is small, the Green's function of the wide band is sharply peaked at $k_F^{\mathrm{eff}}$, which is unchanged from $k_F^{(0)}$. Here the renormalization is weak, and so is the pairing interaction.
For $n=2.45$, for which $\lambda$ is large, $k_F^{\mathrm{eff}}$ is shifted from $k_F^{(0)}$, so that the electron band filling of the renormalized dispersive band is smaller compared to that of the bare one. This means that the renormalized flat band is closer to full-filling (note that the bare flat band is not fully filled even though the chemical potential lies above the flat band due to finite temperature effect). In other words, the electron-electron interaction moves the flat band away from the chemical potential. In  this case, the Green's function of the dispersive band is still relatively sharply peaked at $k_F^{\mathrm{eff}}$, indicating that the self-energy renormalization of this band is not so strong, and hence favorable for superconductivity. On the other hand, when $n=2.2$, where $\lambda$ becomes strongly reduced, the band filling of the renormalized dispersive band increases compared to the non-interacting case because $k_F^{\mathrm{eff}}$ is closer to $k_x=0$ or $2\pi$ compared to $k_F^{(0)}$. This implies that the flat band is closer to the chemical potential, affecting strongly the dispersive band. Consequently, the peak structure of the Green's function of the dispersive band is suppressed, leading to the suppression of superconductivity.

\begin{figure}
\centering
\textbf{(a) $n = 3.4$}
\\
\begin{minipage}{1.0\hsize}
\includegraphics[width=8.5cm]{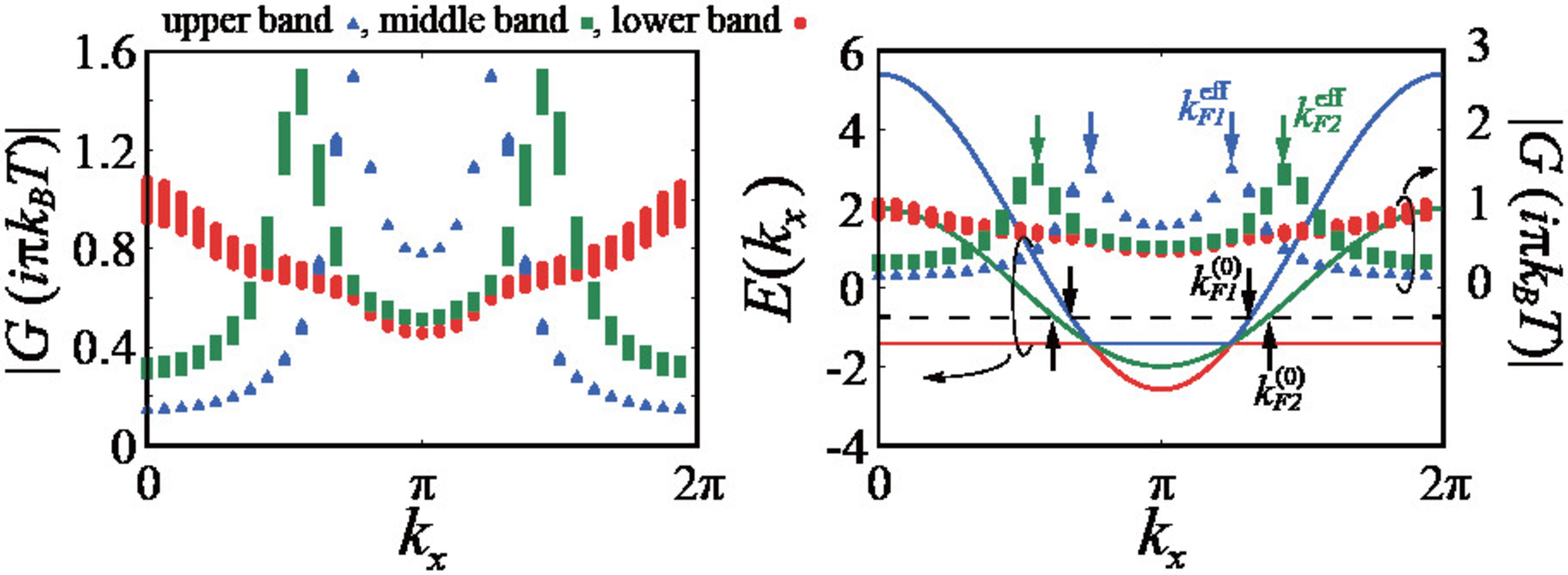}
\end{minipage}
\\
\centering
\textbf{(b) $n = 3.1$}
\\
\begin{minipage}{1.0\hsize}
\includegraphics[width=8.5cm]{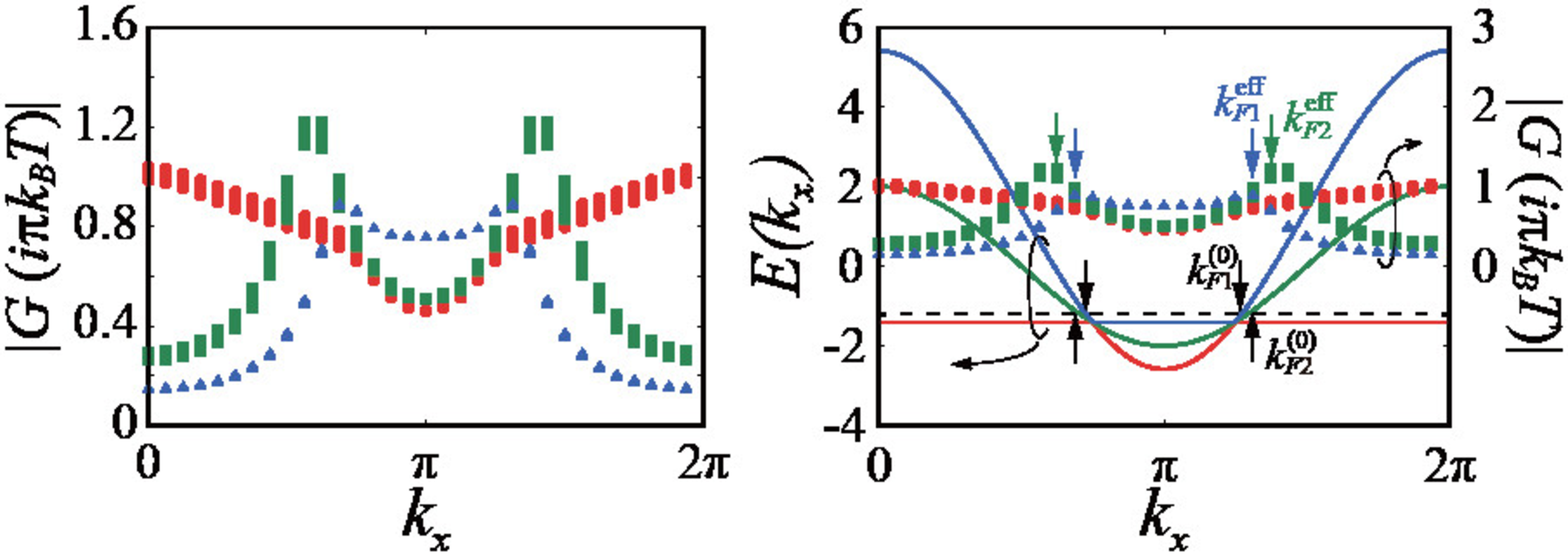}
\end{minipage}
\caption{Green's function for the three-leg ladder plotted against $k_x$  (a)$n=3.4$,(b)$n=3.1$}
\label{fig6}
\end{figure}

\begin{figure}
\includegraphics[width=7.5cm]{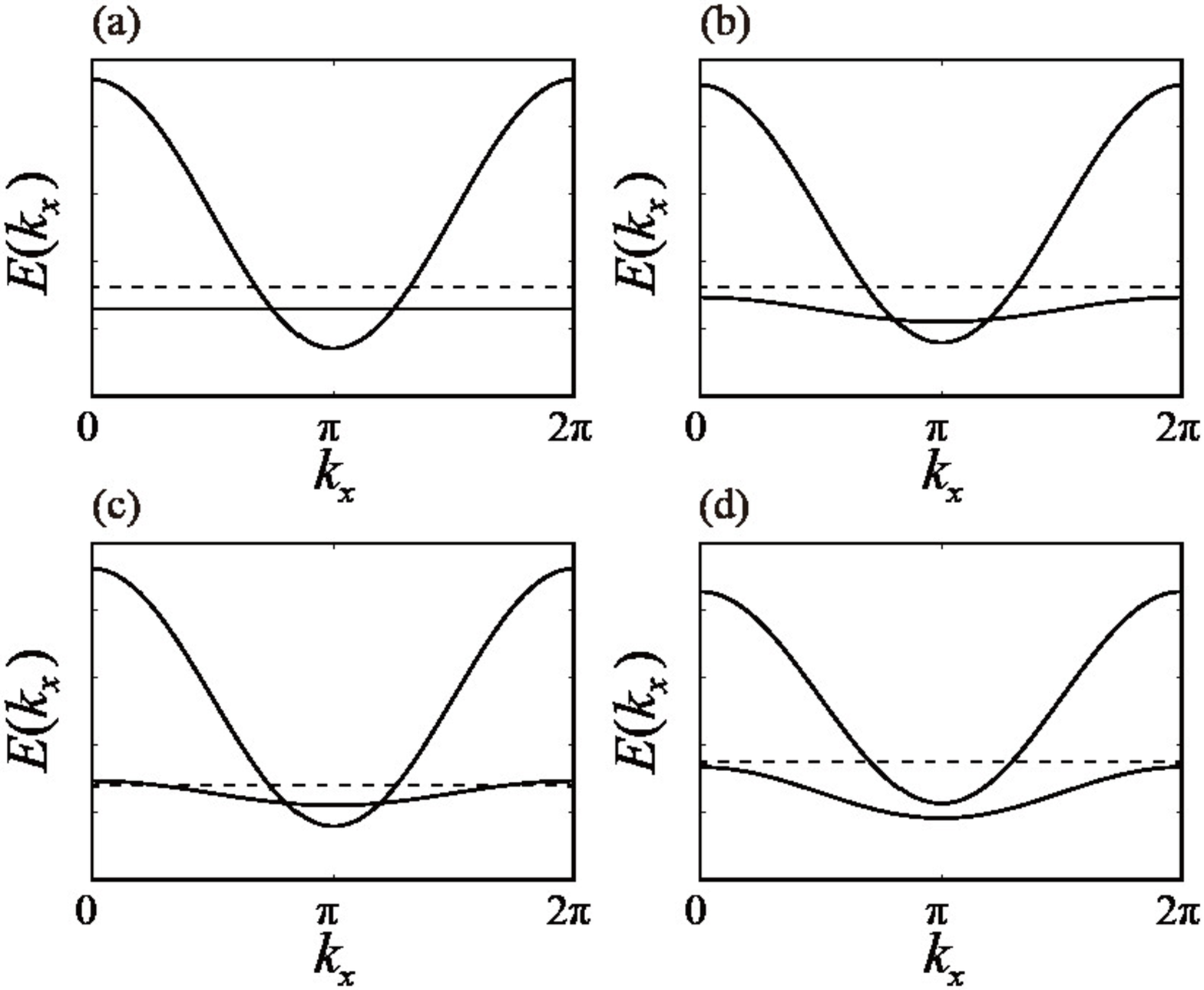}
\caption{Schematic image of the effective band structure of the three-leg ladder, where the two bands effective to superconductivity are extracted. (a) the bare band structure when the narrow band is perfectly flat. (b) when $U$ is introduced in the case of (a) ; here superconductivity is optimized within cases where the bare narrow band is flat (corresponds to $n=3.4$, $t'=1/\sqrt{2}$). (c) when the chemical potential is reduced from the case of (b); superconductivity is suppressed compared to (b) (corresponds to $n=3.1$, $t'=1/\sqrt{2}$). (d) when $t'$ is reduced from the case of (b) (the bare narrow band has finite width); here superconductivity is enhanced compared to (b) (corresponds to $n=3.4$, $t'=0.45$).}
\label{fig7}
\end{figure}

In the three-leg ladder case shown in Fig.\ref{fig6}, the Green's function of the flat band exhibits a significant $k_x$ dependence, indicating that the flat band is effectively ``bent'' by the electron-electron interaction (see the schematic image given in Fig.\ref{fig7}(b))\cite{3leg_flat_comment}. For $n=3.4$, where the eigenvalue $\lambda$ is large, $k_{F1}^{\mathrm{eff}}$ is smaller than $k_{F1}^{(0)}$. As in the case of the two-leg ladder, this indicates that the band filling of the dispersive band is reduced due to the on-site interaction, and hence the band filling of the flat band increases (moves away from the chemical potential), despite the bending of the flat band. By contrast, for $n=3.1$, for which $\lambda$ is suppressed, $k_{F1}^{\mathrm{eff}}$ is shifted from $k_{F1}^{(0)}$ in a manner that the band filling of the renormalized dispersive band {\it increases} compared to the non-interacting case, even though the bare chemical potential still lies well above the bare flat band. Such a shift occurs more strongly  in the third (the other dispersive) band as seen from the comparison between $k_{F2}^{\mathrm{eff}}$ and $k_{F2}^{(0)}$. This means that the flat band is effectively bent to come closer to (or intersect)  the chemical potential near $k_x=0$ (schematic image in Fig.\ref{fig7}(c)). In quasi-one-dimensional systems, even when the band is bent, the density of states at the band edge is large and hence can mimic the effect of the flat band, at least to some extent.  Therefore, the band edge approaching the chemical potential affects the dispersive band, so that the peak structure of the Green's function of the dispersive band is strongly suppressed. This ``bending'' of the flat band, which does not take place in the two-leg ladder, is the reason why superconductivity is suppressed even when the bare chemical potential lies somewhat away from the bare flat band position. 

\begin{figure}
\centering
\textbf{(a) $n = 1.9$}
\\
\begin{minipage}{1.0\hsize}
\includegraphics[width=8.5cm]{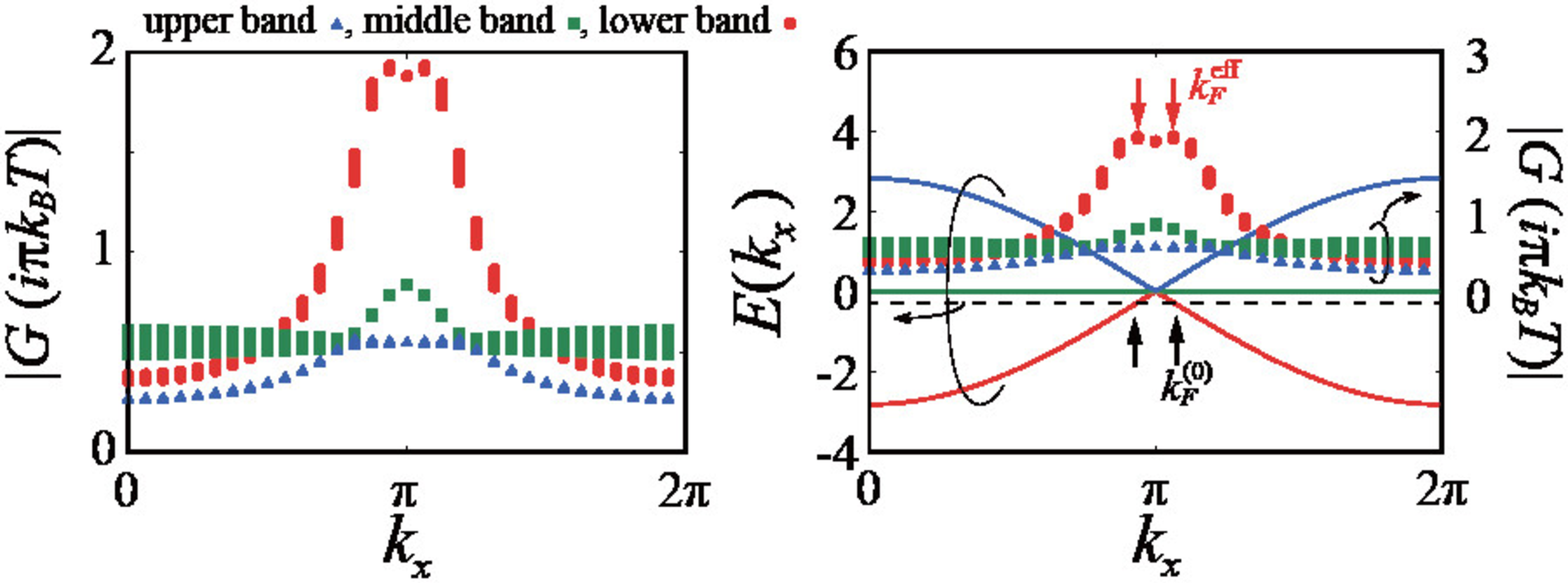}
\end{minipage}
\\
\centering
\textbf{(b) $n = 1.98$}
\\
\begin{minipage}{1.0\hsize}
\includegraphics[width=8.5cm]{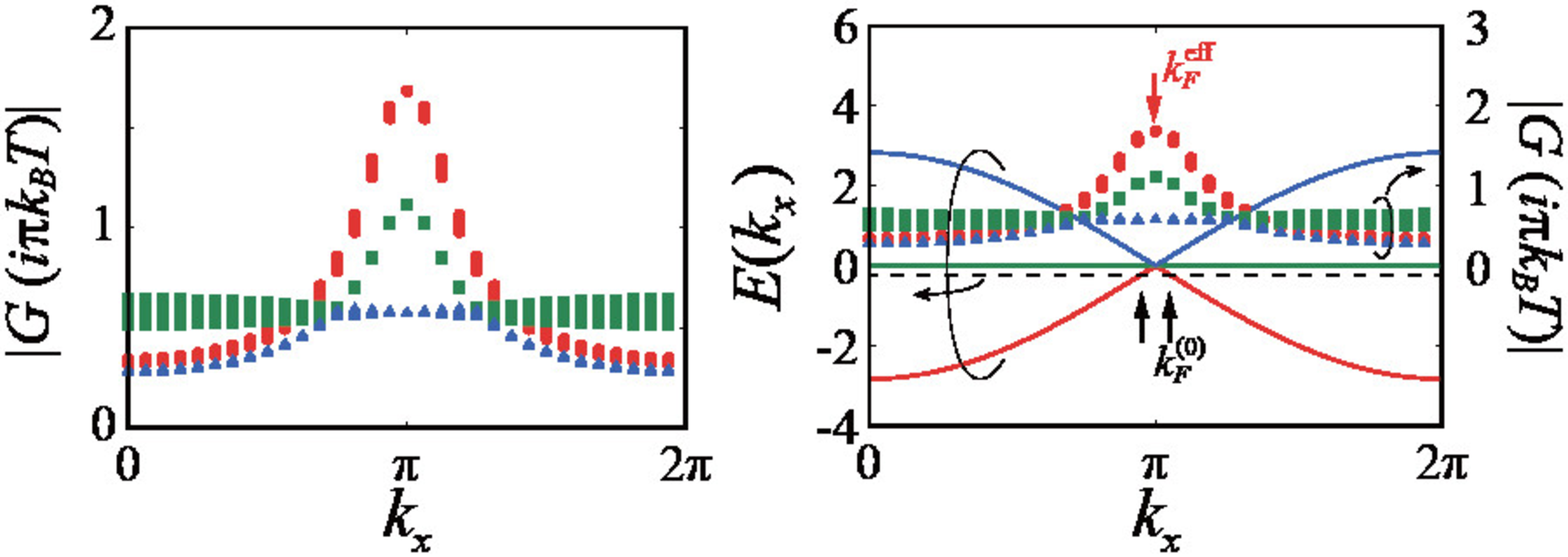}
\end{minipage}
\caption{Green's function for the diamond chain with $t'=0$  plotted against $k_x$ (a)$n=1.9$, (b)$n=1.98$}
\label{fig8}
\end{figure}

As for the diamond chain with $t'=0$, the Green's function of the flat band remains to be $k_x$ independent, and $k_F^{\mathrm{eff}}$ of the lower dispersive band is nearly the same as  $k_F^{(0)}$ for all the band fillings shown in Fig.\ref{fig8}, indicating that the position of the flat band measured from the chemical potential is barely changed by the electron-electron interaction. Hence the situation here is between the two-leg and three-leg cases; due to the interaction, the flat band moves away from the chemical potential in the former, and moves towards it in the latter due to the band bending. 

For the diamond chain with $t'=1$, the situation is complicated. Although $\lambda$ as a function of the chemical potential is peaked near the flat band energy, further calculation results regarding the temperature dependence suggest  that superconductivity actually does not take place in this case, consistent with the conclusion in Ref.\onlinecite{Kobayashi2016_flatband}. We will come back to this point later.

\subsection{Cases with a narrow band with finite dispersion}
\begin{figure}
\includegraphics[width=8.5cm]{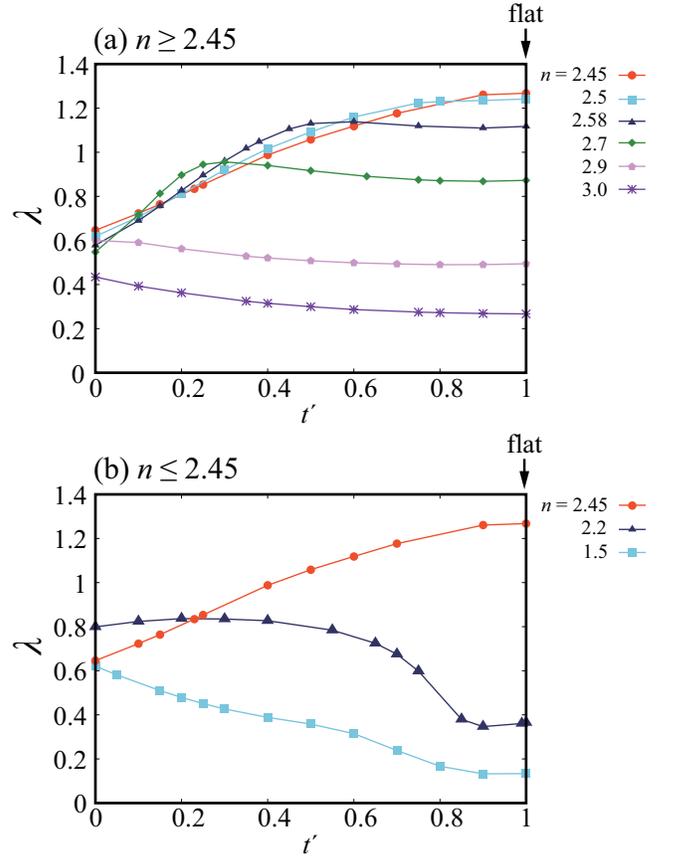}
\caption{$\lambda$ plotted against $t'$ for the two-leg ladder for various band fillings.}
\label{fig9}
\end{figure}

We now move on to the case when the bare narrow band has finite dispersion in the $k_x$ direction. Once again we start with the two-leg ladder as shown in Fig.\ref{fig9}. We reduce $t'$ from 1, which will make the flat band have finite band width. For the band filling of $n=2.45$, where $\lambda$ was the largest for the flat-band case, $\lambda$ monotonically decreases as the dispersion of the narrow band increases. On the other hand, for larger band fillings like $n=2.58$ or $n=2.7$, $\lambda$ first increases and takes its maximum at a certain $t'$. 

\begin{figure}
\includegraphics[width=8.5cm] {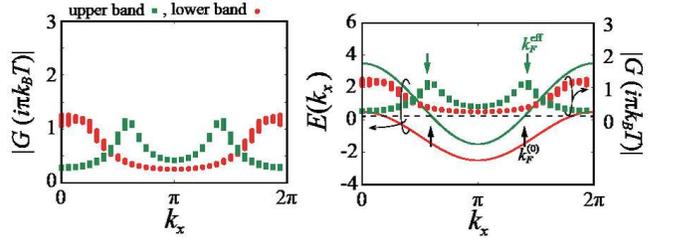}
\caption{Green's function against $k_x$ for the two-leg ladder with $n=2.45$, $t'=0.25$}
\label{fig10}
\end{figure}

These results can again be understood by looking at the renormalized Green's function shown in Fig.\ref{fig10}. For the case of $n=2.45$, when $t'$ is reduced to 0.25, $k_F^{\mathrm{eff}}$ and $k_F^{(0)}$ are nearly the same, implying that the renormalized narrow band intersects the chemical potential just as in the non-interacting case. Consequently, the peak structure of the Green's function of the wide band is suppressed compared to the flat band case of $t'=1$, and hence superconductivity is degraded.

\begin{figure}
\centering
\textbf{(a) $t^\prime = 1$}
\\
\begin{minipage}{1.0\hsize}
\includegraphics[width=8.5cm]{fig5-a.eps}
\end{minipage}
\\
\centering
\textbf{(b) $t^\prime = 0.3$}
\\
\begin{minipage}{1.0\hsize}
\includegraphics[width=8.5cm]{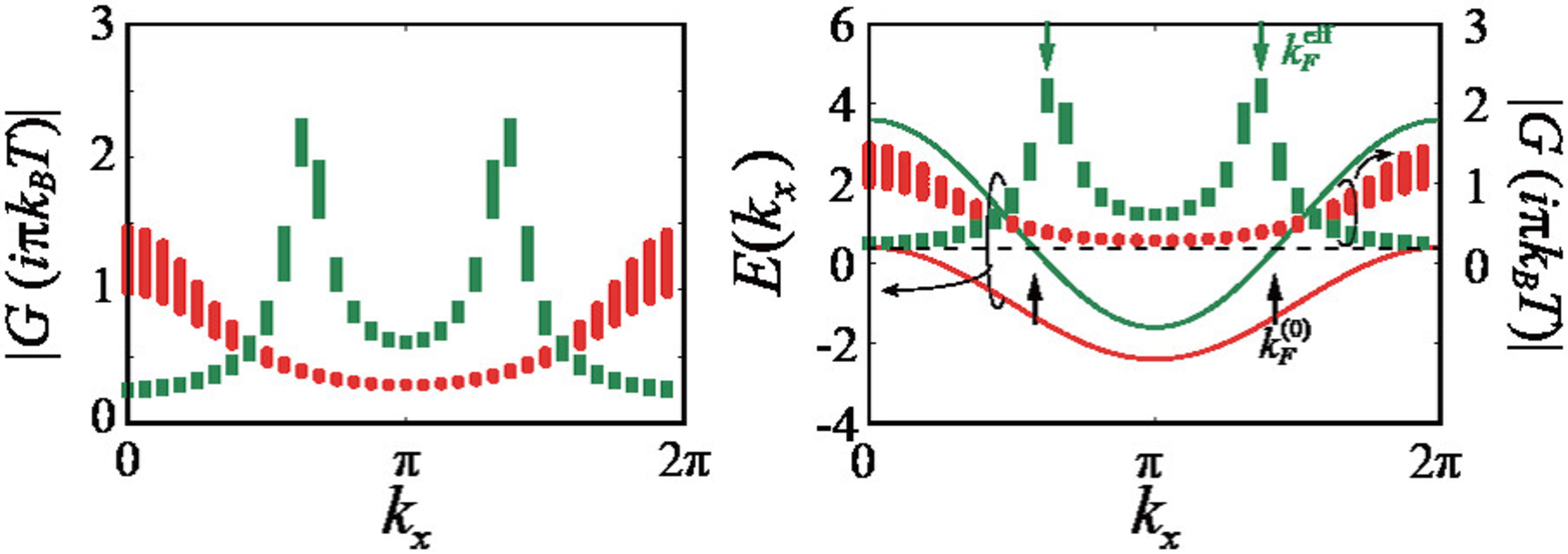}
\end{minipage}
\\
\centering
\textbf{(c) $t^\prime = 0$}
\\
\begin{minipage}{1.0\hsize}
\includegraphics[width=8.5cm]{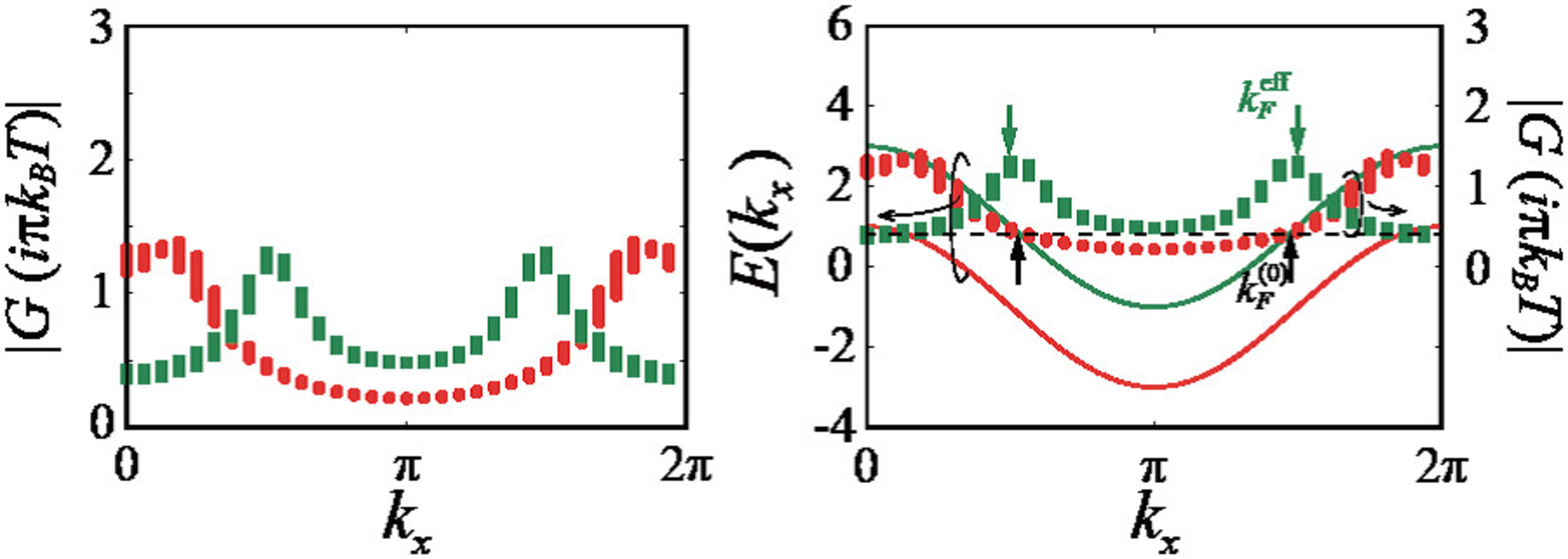}
\end{minipage}
\caption{Green's function against $k_x$ for the two-leg ladder with $n=2.7$ and (a) $t'=1$, (b) $t'=0.3$, or (c) $t'=0$}
\label{fig11}
\end{figure}

On the other hand, when $n=2.7$ (Fig.\ref{fig11}), the flat band lies far away from the chemical potential for $t'=1$, so that the pairing interaction is weak and superconductivity is suppressed,  but reducing $t'$ to 0.3 makes the narrow band edge around $k_x=0$ approach the chemical potential, as can be seen in the enhancement of its Green' function peak around $k_x=0$. This enhances the pairing interaction and hence $\lambda$, since the edge of a quasi-one-dimensional band, due to its large density of states, can play the role of a flat band, as mentioned in section\ref{flatcase}. Too large dispersion (too small $t'$) makes the narrow band intersect the chemical potential, as seen from the fact that the peak of the Green' function of the lower band shifts from $k_x=0$ for $t'=0$. This suppresses superconductivity, once again similar to the case when the chemical potential comes too close to the perfectly flat band.

It is also interesting to study the band width dependence of cases when the chemical potential is too close to the flat band at $t'=1$, so that superconductivity is strongly degraded. In the lower panel of Fig.\ref{fig9}, we display $\lambda$ against $t'$ for $n\leq 2.45$. In contrast to the case of $n=2.45$, $\lambda$ increases as $t'$ is reduced in cases with smaller $n$. This is because the strong renormalization effect of the flat band is reduced by the introduction of the finite band width.

\begin{figure}
\includegraphics[width=7.5cm]{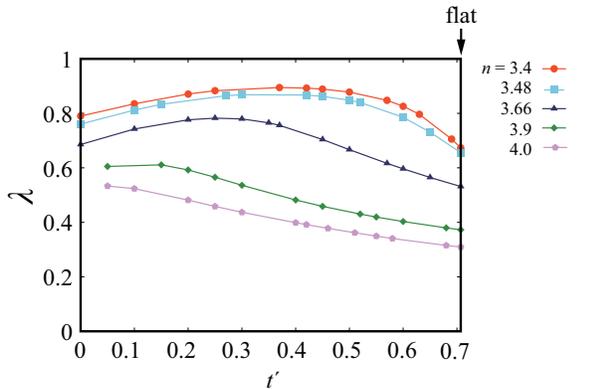}
\caption{$\lambda$ plotted against $t'$ for the three-leg ladder for various band fillings.}
\label{fig12}
\end{figure}

\begin{figure}
\centering
\textbf{(a) $t^\prime = 1/\sqrt{2}$}
\\
\begin{minipage}{1.0\hsize}
\includegraphics[width=8.5cm]{fig6-a.eps}
\end{minipage}
\\
\centering
\textbf{(b) $t^\prime = 0.45$}
\\
\begin{minipage}{1.0\hsize}
\includegraphics[width=8.5cm]{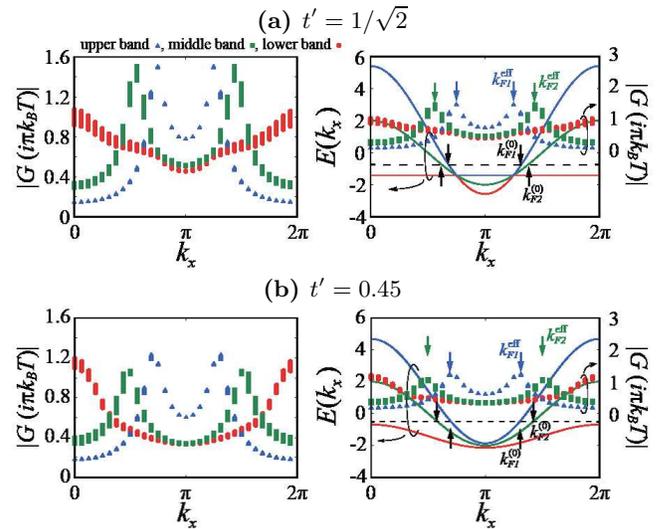}
\end{minipage}
\caption{Green's function against $k_x$ for the three-leg ladder with  $n=3.4$  and (a) $t'=1/\sqrt{2}$ or (b) $t'=0.45$.}
\label{fig13}
\end{figure}

Let us now move on to the three-leg ladder. As shown in Fig.\ref{fig12}, when $t'$ is reduced from its flat-band value $1/\sqrt{2}$, $\lambda$ increases even for the band filling of $n=3.4$, which gives the largest $\lambda$ for the flat-band case. This is in contrast to the two-leg situation, where $\lambda$ monotonically deceases with the reduction of $t'$ for the band filling that gives the largest $\lambda$ for the flat-band case.  This result for the three-leg ladder can again be traced back to the renormalized Green's function shown in Fig.\ref{fig13}. For $t'=0.45$, where $\lambda$ is maximized, it can be seen that the Green's function of the narrow band is peaked around $k_x=0$ showing that the band edge has approached the chemical potential, which makes the pairing interaction stronger. Still, the Green's function of the wide band is not strongly suppressed compared to $t'=1/\sqrt{2}$, which indicates that the renormalization is not strong. This combination of the large pairing interaction and the weak renormalization results in the enhancement of superconductivity with the reduction of $t'$. As mentioned in the previous section, for $t'=1/\sqrt{2}$ and $n=3.4$, the bare chemical potential is positioned relatively far from the flat band, so that there is still some room for $\lambda$ to be enhanced by increasing the band width of the narrow band.  This is in contrast to the case when $t'=1/\sqrt{2}$ is fixed (the bare band is kept flat) and the band filling (the chemical potential) is reduced, where the renormalization of the wide band becomes strong. The difference between the two cases lies in that only the portion of the band around $k_x=0$ approaches the chemical potential by reducing $t'$ and fixing $n$, while the entire flat band approaches it when  $n$ is reduced with a fixed $t'$  (schematic images in Fig.\ref{fig7}(c) and (d)). 

\begin{figure}
\begin{minipage}{0.48\hsize}
\includegraphics[width=4.2cm] {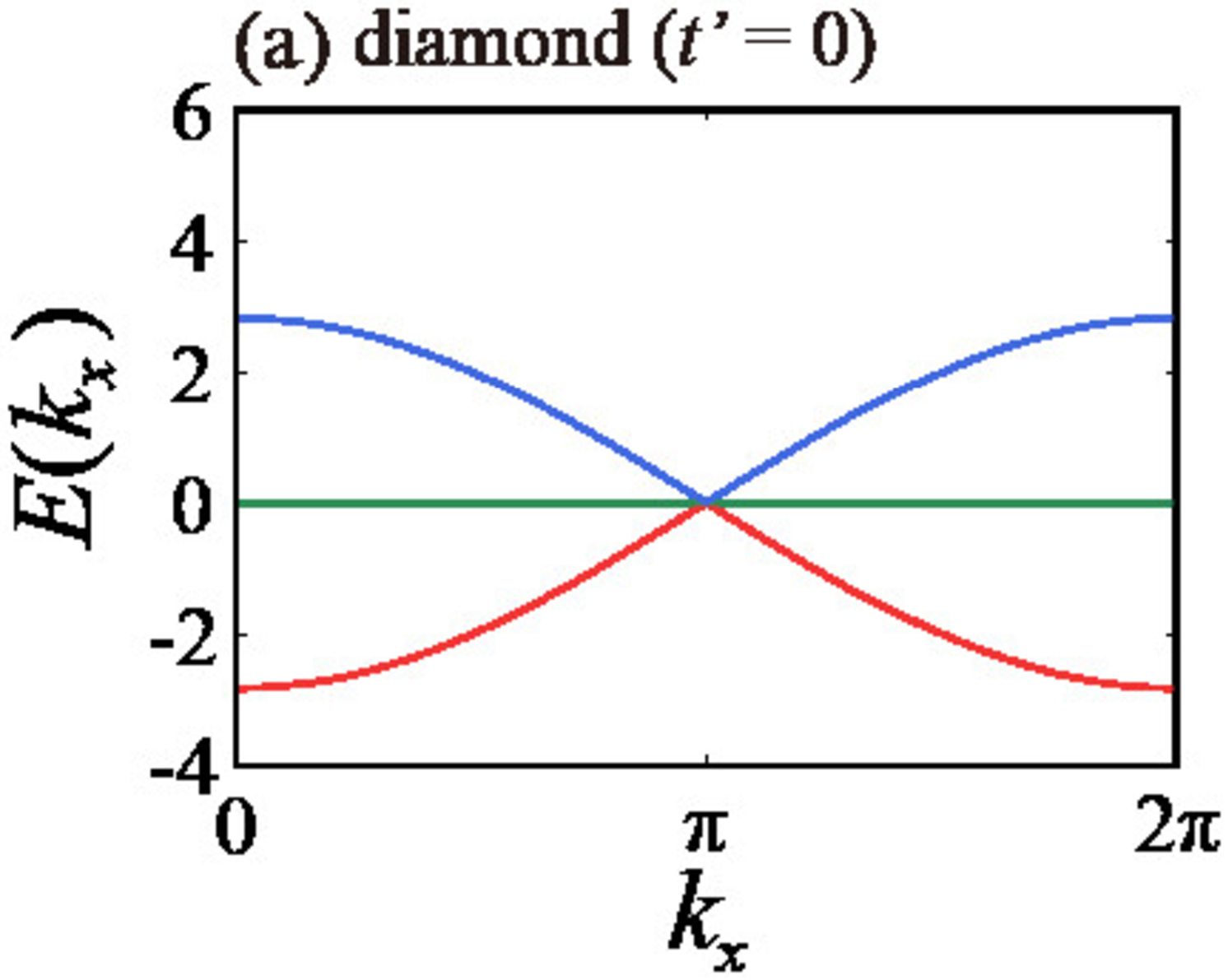}
\end{minipage}
\\
\begin{minipage}{0.48\hsize}
\includegraphics[width=4.2cm] {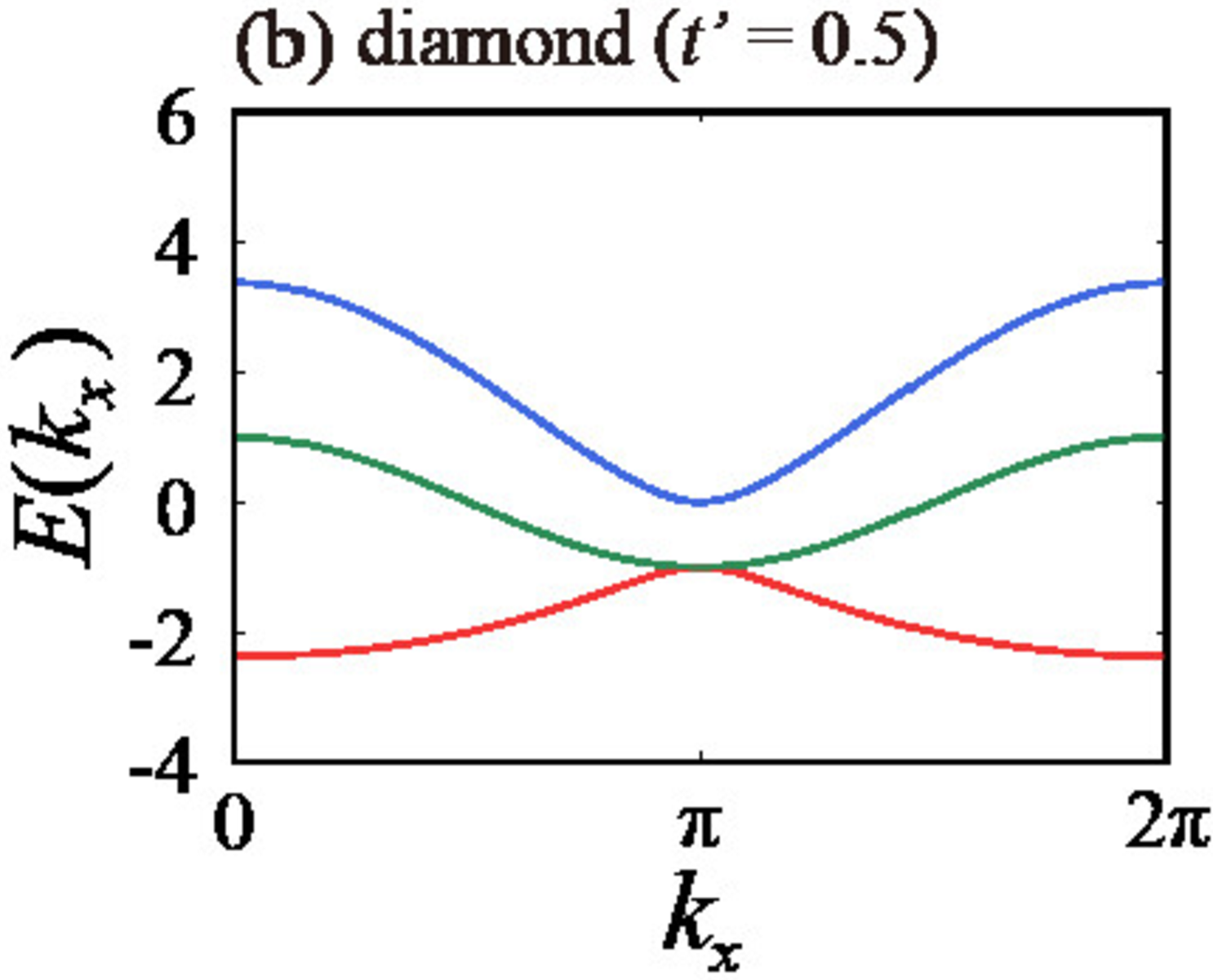}
\end{minipage}
\\
\begin{minipage}{0.48\hsize}
\includegraphics[width=4.2cm] {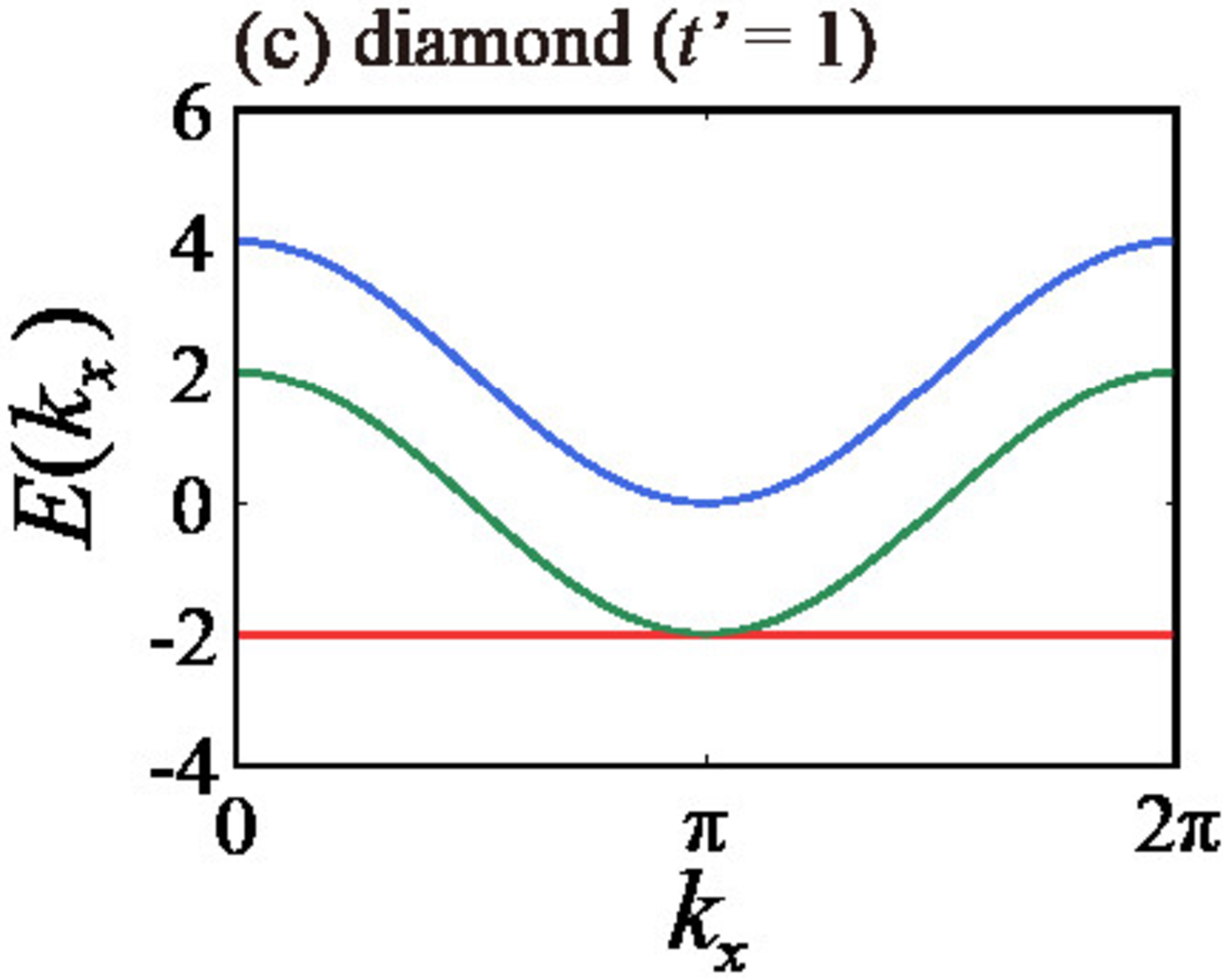}
\end{minipage}
\caption{Evolution of the band structure of the diamond chain with (a) $t'=0$ (b) $t'=0.5$ (c) $t'=1$.}
\label{fig14}
\end{figure}

\begin{figure}
\includegraphics[width=8.5cm]{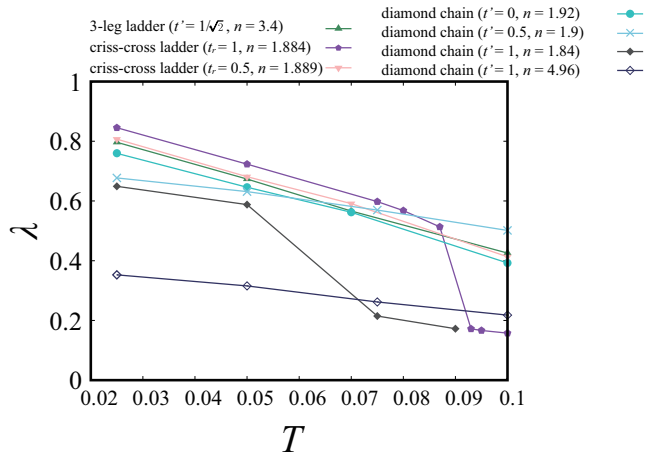}
\caption{$\lambda$ plotted against the temperature for the diamond chain with $t'=0$, $t'=0.5$, or $t'=1$, along with the results for the three-leg ladder with $t'=1/\sqrt{2}$ and those for the criss-cross ladder with $t_r=1$ or 0.5.}
\label{fig15}
\end{figure}

For the diamond chain, we can continuously connect the two flat-band cases by varying $t'$ from 0 to 1. In Fig.\ref{fig14}, we show the evolution of the band dispersion. As can be seen from this evolution, the edge of the lowest and the middle bands have zero group velocity when $t'$ is turned on, so according to our previous discussion, these two bands both play the role of the flat band when the chemical potential is placed in the vicinity of these band edges, while neither of them can act as the wide band. This suggests the possibility that superconductivity actually does not take place at least for large values of $t'$. To investigate this possibility, in Fig.\ref{fig15} we display the temperature dependence of the eigenvalue, fixing the band filling at which the eigenvalue is nearly maximized at $T=0.05$. For $t'=1$, there is indeed a strong tendency of saturation of $\lambda$ upon lowering the temperature, which suggests the absence of superconductivity. Even at $t'=0.5$, the growth of $\lambda$ is slow. These results are consistent with the conclusion of Ref.\onlinecite{Kobayashi2016_flatband} as well as the expectation drawn from the fact that all the bands have zero gradient near the chemical potential for finite values of $t'$, so that none of the bands can act as the wide band. 

In the above context, it is also worth commenting on the peak of $\lambda$ against $n$ in the diamond chain with $t'=1$ around $n=5$  (Fig.\ref{fig3}(b)). This corresponds to the situation when the chemical potential is around 2 (see Fig.\ref{fig14}(c)), where it intersects the wide band and is in the vicinity of the top of the second band\cite{diamond_half_comment}. The peak in $\lambda$ originates from the coexistence of the light mass of the wide band and the large density of states at the narrow band edge. However, the eigenvalue is relatively small, and studying its temperature dependence strongly suggests absence of superconductivity, as shown in Fig.\ref{fig15}. This reconfirms our view that the band filling should not be too far away from half-filling for the present high $T_c$ mechanism to be effective.

\section{CONCLUSION}
Our present study shows that the high $T_c$ pairing mechanism due to coexisting wide and incipient narrow bands works quite generally in quasi-one-dimensional systems. When the chemical potential sits close to the narrow band, the pairing interaction due to interband pair scattering becomes large and is favorable for superconductivity. However, if the chemical potential comes too close to the narrow band, the renormalization effect becomes strong and superconductivity is degraded. In quasi-one-dimensional systems, the density of states at the band edge is (nearly) diverging, so that the band edge itself can play the role of the flat band to some extent, even if the total band width is not so small. Hence, superconductivity can in some cases be optimized by increasing the dispersion of the narrow band so as to make the band edge approach the chemical potential. In the two-leg ladder, superconductivity is sharply optimized when the bare chemical potential sits in the very vicinity of the flat band or the narrow band edge. On the other hand, in the three-leg ladder, superconductivity is optimized when the bare chemical potential lies relatively far from the flat band. The difference between the two cases originates from the bending of the flat band due to electron-electron interaction, which occurs in the latter case. For the diamond chain, the case with $t'=0$ is special in that the wide band has finite gradient at the chemical potential. When $t'$ is finite, the lower two bands have zero gradient at the band edges, so that neither of them can play the role of the wide band when the Fermi level is positioned in the vicinity of the two band edges. Consequently, the calculation result of the temperature dependence suggests that superconductivity does not take place at least when $t'$ is large. 

Here we summarize the condition for the present high $T_c$ pairing mechanism to be effective. The first is that the wide and narrow bands arise from multiple sites within a unit cell, where each site corresponds to a single orbital, namely, the two bands must originate from the same orbital. This is because the origin of the pair scattering is the spin fluctuation, which is strongly enhanced only by the intraorbital on-site repulsion. The second is that the position of the narrow band edge must be in the energy range where the wide band has large gradient. Especially when the narrow band is flat, the flat band has to intersect the wide band. As for this condition,  a recent study provides a general prescription for obtaining such kind of coexisting flat and dispersive bands\cite{Misumi_flatband}.  The third is that the Fermi level must be placed in the vicinity of, but not too close to, the narrow band edge. Superconductivity occurs rather abruptly against the variation of the Fermi level or the temperature (see the abrupt jump of $\lambda$ against $T$ in Fig.\ref{fig15}). In addition, the corresponding band filling should be somewhat close to half-filling; otherwise the electron correlation will be too weak to induce high $T_c$.

As for actual materials in which the present mechanism can be realized, the ladder-type cuprates are candidates, as proposed in Ref.\onlinecite{Kuroki2005_wide-narrow}. There, only the two-leg ladder was considered, but the three-leg ladder cuprate Sr$_2$Cu$_3$O$_5$ can also be a candidate according to the present study. We stress here that although there are no flat bands in the band structure of these materials (they roughly correspond to the case of $t'\sim 0.3$\cite{Muller_SrCu2O3}), the narrow band edge plays the role of the flat band, as was revealed in the present study. However, a drawback of these ladder-type cuprates is that they are notorious for being unable to dope large amount of carriers, especially, electrons. Large amount of electron doping required for realizing high $T_c$ superconductivity is unlikely to be attained by element substitution in bulk ladder cuprates. Recently, two of the present authors proposed another way of realizing the present mechanism by introducing a concept of ``hidden ladders'' in Ruddlesden-Popper compounds\cite{Ogura_Hiddenladder}, where $d_{xz}$ and $d_{yz}$ orbitals form ladders in the $x$ and $y$ directions, respectively. There, materials such as Sr$_3$Mo$_2$O$_7$ and Sr$_3$Cr$_2$O$_7$ have been proposed as good candidates in which the Fermi level is placed in the vicinity of the narrow band.

The present study has been restricted to quasi-one-dimensional systems. It is intriguing to see to what extent the ``wide and incipient narrow band pairing mechanism'' remains valid in higher dimensions. In fact, recent studies on the two-dimensional bilayer Hubbard model show how finite energy spin fluctuations enhances superconductivity\cite{Mishra2016_bilayer,Nakata2017_finite-energy}. In this model, the spin fluctuation arises due to the nesting of the Fermi surfaces of bonding and antibonding bands, but the superconductivity mediated by the spin fluctuation is strongly enhanced when the Fermi surface nesting is degraded to some extent, so that the spin fluctuation has large weight in the finite energy regime. An interesting extension of this study is to make one of the bands, say, the antibonding band, narrow (or even flat). Such a study is underway, which suggests that the incipient-narrow-band pairing mechanism does work in two dimensions at least when the density of states of the narrow band is sufficiently large. This problem will serve as an interesting future study, which will be published elsewhere.

\section*{Acknowledgment}
We appreciate Hideo Aoki,  Masayuki Ochi and Hidetomo Usui for fruitful discussions, and Katsuhiro Suzuki for assistance with the FLEX code. 
Part of the numerical calculations were performed at the Supercomputer Center,
Institute for Solid State Physics, University of Tokyo.
This study has been supported by JSPS KAKENHI (Grants JP26247057 and JP16H04338), a Grant-in-Aid for Scientific Research on Innovative Areas (JP17H05481)
D.O. acknowledges support from Grant-in-Aid for JSPS Research Fellow Grant JP16J01021.

\bibliography{paper18}

\begin{thebibliography}{38}%
\makeatletter
\providecommand \@ifxundefined [1]{%
 \@ifx{#1\undefined}
}%
\providecommand \@ifnum [1]{%
 \ifnum #1\expandafter \@firstoftwo
 \else \expandafter \@secondoftwo
 \fi
}%
\providecommand \@ifx [1]{%
 \ifx #1\expandafter \@firstoftwo
 \else \expandafter \@secondoftwo
 \fi
}%
\providecommand \natexlab [1]{#1}%
\providecommand \enquote  [1]{``#1''}%
\providecommand \bibnamefont  [1]{#1}%
\providecommand \bibfnamefont [1]{#1}%
\providecommand \citenamefont [1]{#1}%
\providecommand \href@noop [0]{\@secondoftwo}%
\providecommand \href [0]{\begingroup \@sanitize@url \@href}%
\providecommand \@href[1]{\@@startlink{#1}\@@href}%
\providecommand \@@href[1]{\endgroup#1\@@endlink}%
\providecommand \@sanitize@url [0]{\catcode `\\12\catcode `\$12\catcode
  `\&12\catcode `\#12\catcode `\^12\catcode `\_12\catcode `\%12\relax}%
\providecommand \@@startlink[1]{}%
\providecommand \@@endlink[0]{}%
\providecommand \url  [0]{\begingroup\@sanitize@url \@url }%
\providecommand \@url [1]{\endgroup\@href {#1}{\urlprefix }}%
\providecommand \urlprefix  [0]{URL }%
\providecommand \Eprint [0]{\href }%
\providecommand \doibase [0]{http://dx.doi.org/}%
\providecommand \selectlanguage [0]{\@gobble}%
\providecommand \bibinfo  [0]{\@secondoftwo}%
\providecommand \bibfield  [0]{\@secondoftwo}%
\providecommand \translation [1]{[#1]}%
\providecommand \BibitemOpen [0]{}%
\providecommand \bibitemStop [0]{}%
\providecommand \bibitemNoStop [0]{.\EOS\space}%
\providecommand \EOS [0]{\spacefactor3000\relax}%
\providecommand \BibitemShut  [1]{\csname bibitem#1\endcsname}%
\let\auto@bib@innerbib\@empty
\bibitem [{\citenamefont {White}\ \emph {et~al.}(1989)\citenamefont {White},
  \citenamefont {Scalapino}, \citenamefont {Sugar}, \citenamefont {Loh},
  \citenamefont {Gubernatis},\ and\ \citenamefont {Scalettar}}]{White_QMC}%
  \BibitemOpen
  \bibfield  {author} {\bibinfo {author} {\bibfnamefont {S.~R.}\ \bibnamefont
  {White}}, \bibinfo {author} {\bibfnamefont {D.~J.}\ \bibnamefont
  {Scalapino}}, \bibinfo {author} {\bibfnamefont {R.~L.}\ \bibnamefont
  {Sugar}}, \bibinfo {author} {\bibfnamefont {E.~Y.}\ \bibnamefont {Loh}},
  \bibinfo {author} {\bibfnamefont {J.~E.}\ \bibnamefont {Gubernatis}}, \ and\
  \bibinfo {author} {\bibfnamefont {R.~T.}\ \bibnamefont {Scalettar}},\
  }\href@noop {} {\bibfield  {journal} {\bibinfo  {journal} {Phys. Rev. B}\
  }\textbf {\bibinfo {volume} {40}},\ \bibinfo {pages} {506} (\bibinfo {year}
  {1989})}\BibitemShut {NoStop}%
\bibitem [{\citenamefont {Kuroki}\ \emph {et~al.}(2005)\citenamefont {Kuroki},
  \citenamefont {Higashida},\ and\ \citenamefont
  {Arita}}]{Kuroki2005_wide-narrow}%
  \BibitemOpen
  \bibfield  {author} {\bibinfo {author} {\bibfnamefont {K.}~\bibnamefont
  {Kuroki}}, \bibinfo {author} {\bibfnamefont {T.}~\bibnamefont {Higashida}}, \
  and\ \bibinfo {author} {\bibfnamefont {R.}~\bibnamefont {Arita}},\ }\href
  {\doibase 10.1103/PhysRevB.72.212509} {\bibfield  {journal} {\bibinfo
  {journal} {Phys. Rev. B}\ }\textbf {\bibinfo {volume} {72}},\ \bibinfo
  {pages} {212509} (\bibinfo {year} {2005})}\BibitemShut {NoStop}%
\bibitem [{\citenamefont {Suhl}\ \emph {et~al.}(1959)\citenamefont {Suhl},
  \citenamefont {Matthias},\ and\ \citenamefont {Walker}}]{Suhl}%
  \BibitemOpen
  \bibfield  {author} {\bibinfo {author} {\bibfnamefont {H.}~\bibnamefont
  {Suhl}}, \bibinfo {author} {\bibfnamefont {B.~T.}\ \bibnamefont {Matthias}},
  \ and\ \bibinfo {author} {\bibfnamefont {L.~R.}\ \bibnamefont {Walker}},\
  }\href@noop {} {\bibfield  {journal} {\bibinfo  {journal} {Phys. Rev. Lett.}\
  }\textbf {\bibinfo {volume} {3}},\ \bibinfo {pages} {552} (\bibinfo {year}
  {1959})}\BibitemShut {NoStop}%
\bibitem [{\citenamefont {Kondo}(1963)}]{Kondo}%
  \BibitemOpen
  \bibfield  {author} {\bibinfo {author} {\bibfnamefont {J.}~\bibnamefont
  {Kondo}},\ }\href@noop {} {\bibfield  {journal} {\bibinfo  {journal} {Prog.
  Theor. Phys.}\ }\textbf {\bibinfo {volume} {29}},\ \bibinfo {pages} {1}
  (\bibinfo {year} {1963})}\BibitemShut {NoStop}%
\bibitem [{\citenamefont {Hirschfeld}\ \emph {et~al.}(2011)\citenamefont
  {Hirschfeld}, \citenamefont {Korshunov},\ and\ \citenamefont
  {Mazin}}]{Hirschfeld2011_incipient}%
  \BibitemOpen
  \bibfield  {author} {\bibinfo {author} {\bibfnamefont {P.~J.}\ \bibnamefont
  {Hirschfeld}}, \bibinfo {author} {\bibfnamefont {M.~M.}\ \bibnamefont
  {Korshunov}}, \ and\ \bibinfo {author} {\bibfnamefont {I.~I.}\ \bibnamefont
  {Mazin}},\ }\href {http://stacks.iop.org/0034-4885/74/i=12/a=124508}
  {\bibfield  {journal} {\bibinfo  {journal} {Reports on Progress in Physics}\
  }\textbf {\bibinfo {volume} {74}},\ \bibinfo {pages} {124508} (\bibinfo
  {year} {2011})}\BibitemShut {NoStop}%
\bibitem [{\citenamefont {Miao}\ \emph {et~al.}(2015)\citenamefont {Miao},
  \citenamefont {Qian}, \citenamefont {Shi}, \citenamefont {Richard},
  \citenamefont {Kim}, \citenamefont {Hoesch}, \citenamefont {Xing},
  \citenamefont {Wang}, \citenamefont {Jin}, \citenamefont {Hu} \emph
  {et~al.}}]{Miao2015_ironbased111FS}%
  \BibitemOpen
  \bibfield  {author} {\bibinfo {author} {\bibfnamefont {H.}~\bibnamefont
  {Miao}}, \bibinfo {author} {\bibfnamefont {T.}~\bibnamefont {Qian}}, \bibinfo
  {author} {\bibfnamefont {X.}~\bibnamefont {Shi}}, \bibinfo {author}
  {\bibfnamefont {P.}~\bibnamefont {Richard}}, \bibinfo {author} {\bibfnamefont
  {T.}~\bibnamefont {Kim}}, \bibinfo {author} {\bibfnamefont {M.}~\bibnamefont
  {Hoesch}}, \bibinfo {author} {\bibfnamefont {L.}~\bibnamefont {Xing}},
  \bibinfo {author} {\bibfnamefont {X.-C.}\ \bibnamefont {Wang}}, \bibinfo
  {author} {\bibfnamefont {C.-Q.}\ \bibnamefont {Jin}}, \bibinfo {author}
  {\bibfnamefont {J.-P.}\ \bibnamefont {Hu}},  \emph {et~al.},\ }\href@noop {}
  {\bibfield  {journal} {\bibinfo  {journal} {Nat. Commun.}\ }\textbf {\bibinfo
  {volume} {6}},\ \bibinfo {pages} {6056} (\bibinfo {year} {2015})}\BibitemShut
  {NoStop}%
\bibitem [{\citenamefont {Wang}\ \emph {et~al.}(2011)\citenamefont {Wang},
  \citenamefont {Yang}, \citenamefont {Gao}, \citenamefont {Lu}, \citenamefont
  {Xiang},\ and\ \citenamefont {Lee}}]{Wang2011_122FRG}%
  \BibitemOpen
  \bibfield  {author} {\bibinfo {author} {\bibfnamefont {F.}~\bibnamefont
  {Wang}}, \bibinfo {author} {\bibfnamefont {F.}~\bibnamefont {Yang}}, \bibinfo
  {author} {\bibfnamefont {M.}~\bibnamefont {Gao}}, \bibinfo {author}
  {\bibfnamefont {Z.-Y.}\ \bibnamefont {Lu}}, \bibinfo {author} {\bibfnamefont
  {T.}~\bibnamefont {Xiang}}, \ and\ \bibinfo {author} {\bibfnamefont {D.-H.}\
  \bibnamefont {Lee}},\ }\href {http://stacks.iop.org/0295-5075/93/i=5/a=57003}
  {\bibfield  {journal} {\bibinfo  {journal} {EPL (Europhysics Letters)}\
  }\textbf {\bibinfo {volume} {93}},\ \bibinfo {pages} {57003} (\bibinfo {year}
  {2011})}\BibitemShut {NoStop}%
\bibitem [{\citenamefont {Bang}(2014)}]{Bang2014_122shadowgap}%
  \BibitemOpen
  \bibfield  {author} {\bibinfo {author} {\bibfnamefont {Y.}~\bibnamefont
  {Bang}},\ }\href {http://stacks.iop.org/1367-2630/16/i=2/a=023029} {\bibfield
   {journal} {\bibinfo  {journal} {New Journal of Physics}\ }\textbf {\bibinfo
  {volume} {16}},\ \bibinfo {pages} {023029} (\bibinfo {year}
  {2014})}\BibitemShut {NoStop}%
\bibitem [{\citenamefont {Chen}\ \emph {et~al.}(2015)\citenamefont {Chen},
  \citenamefont {Maiti}, \citenamefont {Linscheid},\ and\ \citenamefont
  {Hirschfeld}}]{Chen2015_incipient}%
  \BibitemOpen
  \bibfield  {author} {\bibinfo {author} {\bibfnamefont {X.}~\bibnamefont
  {Chen}}, \bibinfo {author} {\bibfnamefont {S.}~\bibnamefont {Maiti}},
  \bibinfo {author} {\bibfnamefont {A.}~\bibnamefont {Linscheid}}, \ and\
  \bibinfo {author} {\bibfnamefont {P.~J.}\ \bibnamefont {Hirschfeld}},\ }\href
  {\doibase 10.1103/PhysRevB.92.224514} {\bibfield  {journal} {\bibinfo
  {journal} {Phys. Rev. B}\ }\textbf {\bibinfo {volume} {92}},\ \bibinfo
  {pages} {224514} (\bibinfo {year} {2015})}\BibitemShut {NoStop}%
\bibitem [{\citenamefont {Bang}(2016)}]{Bang2016_dynamicaltuning}%
  \BibitemOpen
  \bibfield  {author} {\bibinfo {author} {\bibfnamefont {Y.}~\bibnamefont
  {Bang}},\ }\href {http://stacks.iop.org/1367-2630/18/i=11/a=113054}
  {\bibfield  {journal} {\bibinfo  {journal} {New Journal of Physics}\ }\textbf
  {\bibinfo {volume} {18}},\ \bibinfo {pages} {113054} (\bibinfo {year}
  {2016})}\BibitemShut {NoStop}%
\bibitem [{\citenamefont {Guo}\ \emph {et~al.}(2010)\citenamefont {Guo},
  \citenamefont {Jin}, \citenamefont {Wang}, \citenamefont {Wang},
  \citenamefont {Zhu}, \citenamefont {Zhou}, \citenamefont {He},\ and\
  \citenamefont {Chen}}]{Guo2010_ironbased122}%
  \BibitemOpen
  \bibfield  {author} {\bibinfo {author} {\bibfnamefont {J.}~\bibnamefont
  {Guo}}, \bibinfo {author} {\bibfnamefont {S.}~\bibnamefont {Jin}}, \bibinfo
  {author} {\bibfnamefont {G.}~\bibnamefont {Wang}}, \bibinfo {author}
  {\bibfnamefont {S.}~\bibnamefont {Wang}}, \bibinfo {author} {\bibfnamefont
  {K.}~\bibnamefont {Zhu}}, \bibinfo {author} {\bibfnamefont {T.}~\bibnamefont
  {Zhou}}, \bibinfo {author} {\bibfnamefont {M.}~\bibnamefont {He}}, \ and\
  \bibinfo {author} {\bibfnamefont {X.}~\bibnamefont {Chen}},\ }\href {\doibase
  10.1103/PhysRevB.82.180520} {\bibfield  {journal} {\bibinfo  {journal} {Phys.
  Rev. B}\ }\textbf {\bibinfo {volume} {82}},\ \bibinfo {pages} {180520}
  (\bibinfo {year} {2010})}\BibitemShut {NoStop}%
\bibitem [{\citenamefont {Qian}\ \emph {et~al.}(2011)\citenamefont {Qian},
  \citenamefont {Wang}, \citenamefont {Jin}, \citenamefont {Zhang},
  \citenamefont {Richard}, \citenamefont {Xu}, \citenamefont {Dai},
  \citenamefont {Fang}, \citenamefont {Guo}, \citenamefont {Chen},\ and\
  \citenamefont {Ding}}]{Qian2011_ironbased122FS}%
  \BibitemOpen
  \bibfield  {author} {\bibinfo {author} {\bibfnamefont {T.}~\bibnamefont
  {Qian}}, \bibinfo {author} {\bibfnamefont {X.-P.}\ \bibnamefont {Wang}},
  \bibinfo {author} {\bibfnamefont {W.-C.}\ \bibnamefont {Jin}}, \bibinfo
  {author} {\bibfnamefont {P.}~\bibnamefont {Zhang}}, \bibinfo {author}
  {\bibfnamefont {P.}~\bibnamefont {Richard}}, \bibinfo {author} {\bibfnamefont
  {G.}~\bibnamefont {Xu}}, \bibinfo {author} {\bibfnamefont {X.}~\bibnamefont
  {Dai}}, \bibinfo {author} {\bibfnamefont {Z.}~\bibnamefont {Fang}}, \bibinfo
  {author} {\bibfnamefont {J.-G.}\ \bibnamefont {Guo}}, \bibinfo {author}
  {\bibfnamefont {X.-L.}\ \bibnamefont {Chen}}, \ and\ \bibinfo {author}
  {\bibfnamefont {H.}~\bibnamefont {Ding}},\ }\href {\doibase
  10.1103/PhysRevLett.106.187001} {\bibfield  {journal} {\bibinfo  {journal}
  {Phys. Rev. Lett.}\ }\textbf {\bibinfo {volume} {106}},\ \bibinfo {pages}
  {187001} (\bibinfo {year} {2011})}\BibitemShut {NoStop}%
\bibitem [{\citenamefont {Qing-Yan}\ \emph {et~al.}(2012)\citenamefont
  {Qing-Yan}, \citenamefont {Zhi}, \citenamefont {Wen-Hao}, \citenamefont
  {Zuo-Cheng}, \citenamefont {Jin-Song}, \citenamefont {Wei}, \citenamefont
  {Hao}, \citenamefont {Yun-Bo}, \citenamefont {Peng}, \citenamefont {Kai},
  \citenamefont {Jing}, \citenamefont {Can-Li}, \citenamefont {Ke},
  \citenamefont {Jin-Feng}, \citenamefont {Shuai-Hua}, \citenamefont {Ya-Yu},
  \citenamefont {Li-Li}, \citenamefont {Xi}, \citenamefont {Xu-Cun},\ and\
  \citenamefont {Qi-Kun}}]{Wang2012_ironbased11-STO}%
  \BibitemOpen
  \bibfield  {author} {\bibinfo {author} {\bibfnamefont {W.}~\bibnamefont
  {Qing-Yan}}, \bibinfo {author} {\bibfnamefont {L.}~\bibnamefont {Zhi}},
  \bibinfo {author} {\bibfnamefont {Z.}~\bibnamefont {Wen-Hao}}, \bibinfo
  {author} {\bibfnamefont {Z.}~\bibnamefont {Zuo-Cheng}}, \bibinfo {author}
  {\bibfnamefont {Z.}~\bibnamefont {Jin-Song}}, \bibinfo {author}
  {\bibfnamefont {L.}~\bibnamefont {Wei}}, \bibinfo {author} {\bibfnamefont
  {D.}~\bibnamefont {Hao}}, \bibinfo {author} {\bibfnamefont {O.}~\bibnamefont
  {Yun-Bo}}, \bibinfo {author} {\bibfnamefont {D.}~\bibnamefont {Peng}},
  \bibinfo {author} {\bibfnamefont {C.}~\bibnamefont {Kai}}, \bibinfo {author}
  {\bibfnamefont {W.}~\bibnamefont {Jing}}, \bibinfo {author} {\bibfnamefont
  {S.}~\bibnamefont {Can-Li}}, \bibinfo {author} {\bibfnamefont
  {H.}~\bibnamefont {Ke}}, \bibinfo {author} {\bibfnamefont {J.}~\bibnamefont
  {Jin-Feng}}, \bibinfo {author} {\bibfnamefont {J.}~\bibnamefont {Shuai-Hua}},
  \bibinfo {author} {\bibfnamefont {W.}~\bibnamefont {Ya-Yu}}, \bibinfo
  {author} {\bibfnamefont {W.}~\bibnamefont {Li-Li}}, \bibinfo {author}
  {\bibfnamefont {C.}~\bibnamefont {Xi}}, \bibinfo {author} {\bibfnamefont
  {M.}~\bibnamefont {Xu-Cun}}, \ and\ \bibinfo {author} {\bibfnamefont
  {X.}~\bibnamefont {Qi-Kun}},\ }\href
  {http://stacks.iop.org/0256-307X/29/i=3/a=037402} {\bibfield  {journal}
  {\bibinfo  {journal} {Chinese Physics Letters}\ }\textbf {\bibinfo {volume}
  {29}},\ \bibinfo {pages} {037402} (\bibinfo {year} {2012})}\BibitemShut
  {NoStop}%
\bibitem [{\citenamefont {Tan}\ \emph {et~al.}(2013)\citenamefont {Tan},
  \citenamefont {Zhang}, \citenamefont {Xia}, \citenamefont {Ye}, \citenamefont
  {Chen}, \citenamefont {Xie}, \citenamefont {Peng}, \citenamefont {Xu},
  \citenamefont {Fan}, \citenamefont {Xu} \emph
  {et~al.}}]{Tan2013_ironbased11-STO}%
  \BibitemOpen
  \bibfield  {author} {\bibinfo {author} {\bibfnamefont {S.}~\bibnamefont
  {Tan}}, \bibinfo {author} {\bibfnamefont {Y.}~\bibnamefont {Zhang}}, \bibinfo
  {author} {\bibfnamefont {M.}~\bibnamefont {Xia}}, \bibinfo {author}
  {\bibfnamefont {Z.}~\bibnamefont {Ye}}, \bibinfo {author} {\bibfnamefont
  {F.}~\bibnamefont {Chen}}, \bibinfo {author} {\bibfnamefont {X.}~\bibnamefont
  {Xie}}, \bibinfo {author} {\bibfnamefont {R.}~\bibnamefont {Peng}}, \bibinfo
  {author} {\bibfnamefont {D.}~\bibnamefont {Xu}}, \bibinfo {author}
  {\bibfnamefont {Q.}~\bibnamefont {Fan}}, \bibinfo {author} {\bibfnamefont
  {H.}~\bibnamefont {Xu}},  \emph {et~al.},\ }\href@noop {} {\bibfield
  {journal} {\bibinfo  {journal} {Nat. Mater.}\ }\textbf {\bibinfo {volume}
  {12}},\ \bibinfo {pages} {634} (\bibinfo {year} {2013})}\BibitemShut
  {NoStop}%
\bibitem [{\citenamefont {Niu}\ \emph {et~al.}(2015)\citenamefont {Niu},
  \citenamefont {Peng}, \citenamefont {Xu}, \citenamefont {Yan}, \citenamefont
  {Jiang}, \citenamefont {Xu}, \citenamefont {Yu}, \citenamefont {Song},
  \citenamefont {Huang}, \citenamefont {Wang}, \citenamefont {Xie},
  \citenamefont {Lu}, \citenamefont {Wang}, \citenamefont {Chen}, \citenamefont
  {Sun},\ and\ \citenamefont {Feng}}]{Niu2015_ironbased11}%
  \BibitemOpen
  \bibfield  {author} {\bibinfo {author} {\bibfnamefont {X.~H.}\ \bibnamefont
  {Niu}}, \bibinfo {author} {\bibfnamefont {R.}~\bibnamefont {Peng}}, \bibinfo
  {author} {\bibfnamefont {H.~C.}\ \bibnamefont {Xu}}, \bibinfo {author}
  {\bibfnamefont {Y.~J.}\ \bibnamefont {Yan}}, \bibinfo {author} {\bibfnamefont
  {J.}~\bibnamefont {Jiang}}, \bibinfo {author} {\bibfnamefont {D.~F.}\
  \bibnamefont {Xu}}, \bibinfo {author} {\bibfnamefont {T.~L.}\ \bibnamefont
  {Yu}}, \bibinfo {author} {\bibfnamefont {Q.}~\bibnamefont {Song}}, \bibinfo
  {author} {\bibfnamefont {Z.~C.}\ \bibnamefont {Huang}}, \bibinfo {author}
  {\bibfnamefont {Y.~X.}\ \bibnamefont {Wang}}, \bibinfo {author}
  {\bibfnamefont {B.~P.}\ \bibnamefont {Xie}}, \bibinfo {author} {\bibfnamefont
  {X.~F.}\ \bibnamefont {Lu}}, \bibinfo {author} {\bibfnamefont {N.~Z.}\
  \bibnamefont {Wang}}, \bibinfo {author} {\bibfnamefont {X.~H.}\ \bibnamefont
  {Chen}}, \bibinfo {author} {\bibfnamefont {Z.}~\bibnamefont {Sun}}, \ and\
  \bibinfo {author} {\bibfnamefont {D.~L.}\ \bibnamefont {Feng}},\ }\href
  {\doibase 10.1103/PhysRevB.92.060504} {\bibfield  {journal} {\bibinfo
  {journal} {Phys. Rev. B}\ }\textbf {\bibinfo {volume} {92}},\ \bibinfo
  {pages} {060504} (\bibinfo {year} {2015})}\BibitemShut {NoStop}%
\bibitem [{\citenamefont {Mishra}\ \emph {et~al.}(2016)\citenamefont {Mishra},
  \citenamefont {Scalapino},\ and\ \citenamefont {Maier}}]{Mishra2016_bilayer}%
  \BibitemOpen
  \bibfield  {author} {\bibinfo {author} {\bibfnamefont {V.}~\bibnamefont
  {Mishra}}, \bibinfo {author} {\bibfnamefont {D.~J.}\ \bibnamefont
  {Scalapino}}, \ and\ \bibinfo {author} {\bibfnamefont {T.}~\bibnamefont
  {Maier}},\ }\href@noop {} {\bibfield  {journal} {\bibinfo  {journal}
  {Scientific reports}\ }\textbf {\bibinfo {volume} {6}},\ \bibinfo {pages}
  {32078} (\bibinfo {year} {2016})}\BibitemShut {NoStop}%
\bibitem [{\citenamefont {Dagotto}\ \emph {et~al.}(1992)\citenamefont
  {Dagotto}, \citenamefont {Riera},\ and\ \citenamefont
  {Scalapino}}]{Dagotto1992_ladder}%
  \BibitemOpen
  \bibfield  {author} {\bibinfo {author} {\bibfnamefont {E.}~\bibnamefont
  {Dagotto}}, \bibinfo {author} {\bibfnamefont {J.}~\bibnamefont {Riera}}, \
  and\ \bibinfo {author} {\bibfnamefont {D.}~\bibnamefont {Scalapino}},\ }\href
  {\doibase 10.1103/PhysRevB.45.5744} {\bibfield  {journal} {\bibinfo
  {journal} {Phys. Rev. B}\ }\textbf {\bibinfo {volume} {45}},\ \bibinfo
  {pages} {5744(R)} (\bibinfo {year} {1992})}\BibitemShut {NoStop}%
\bibitem [{\citenamefont {Rice}\ \emph {et~al.}(1993)\citenamefont {Rice},
  \citenamefont {Gopalan},\ and\ \citenamefont {Sigrist}}]{Rice1993_ladder}%
  \BibitemOpen
  \bibfield  {author} {\bibinfo {author} {\bibfnamefont {T.~M.}\ \bibnamefont
  {Rice}}, \bibinfo {author} {\bibfnamefont {S.}~\bibnamefont {Gopalan}}, \
  and\ \bibinfo {author} {\bibfnamefont {M.}~\bibnamefont {Sigrist}},\ }\href
  {http://stacks.iop.org/0295-5075/23/i=6/a=011} {\bibfield  {journal}
  {\bibinfo  {journal} {EPL (Europhysics Letters)}\ }\textbf {\bibinfo {volume}
  {23}},\ \bibinfo {pages} {445} (\bibinfo {year} {1993})}\BibitemShut
  {NoStop}%
\bibitem [{\citenamefont {Dagotto}\ and\ \citenamefont
  {Rice}(1996)}]{Dagotto-Rice1996_ladder}%
  \BibitemOpen
  \bibfield  {author} {\bibinfo {author} {\bibfnamefont {E.}~\bibnamefont
  {Dagotto}}\ and\ \bibinfo {author} {\bibfnamefont {T.~M.}\ \bibnamefont
  {Rice}},\ }\href {\doibase 10.1126/science.271.5249.618} {\bibfield
  {journal} {\bibinfo  {journal} {Science}\ }\textbf {\bibinfo {volume}
  {271}},\ \bibinfo {pages} {618} (\bibinfo {year} {1996})}\BibitemShut
  {NoStop}%
\bibitem [{\citenamefont {Nagata}\ \emph {et~al.}(1998)\citenamefont {Nagata},
  \citenamefont {Uehara}, \citenamefont {Goto}, \citenamefont {Akimitsu},
  \citenamefont {Motoyama}, \citenamefont {Eisaki}, \citenamefont {Uchida},
  \citenamefont {Takahashi}, \citenamefont {Nakanishi},\ and\ \citenamefont
  {M\^ori}}]{Nagata1998_14-24-41}%
  \BibitemOpen
  \bibfield  {author} {\bibinfo {author} {\bibfnamefont {T.}~\bibnamefont
  {Nagata}}, \bibinfo {author} {\bibfnamefont {M.}~\bibnamefont {Uehara}},
  \bibinfo {author} {\bibfnamefont {J.}~\bibnamefont {Goto}}, \bibinfo {author}
  {\bibfnamefont {J.}~\bibnamefont {Akimitsu}}, \bibinfo {author}
  {\bibfnamefont {N.}~\bibnamefont {Motoyama}}, \bibinfo {author}
  {\bibfnamefont {H.}~\bibnamefont {Eisaki}}, \bibinfo {author} {\bibfnamefont
  {S.}~\bibnamefont {Uchida}}, \bibinfo {author} {\bibfnamefont
  {H.}~\bibnamefont {Takahashi}}, \bibinfo {author} {\bibfnamefont
  {T.}~\bibnamefont {Nakanishi}}, \ and\ \bibinfo {author} {\bibfnamefont
  {N.}~\bibnamefont {M\^ori}},\ }\href {\doibase 10.1103/PhysRevLett.81.1090}
  {\bibfield  {journal} {\bibinfo  {journal} {Phys. Rev. Lett.}\ }\textbf
  {\bibinfo {volume} {81}},\ \bibinfo {pages} {1090} (\bibinfo {year}
  {1998})}\BibitemShut {NoStop}%
\bibitem [{\citenamefont {Kobayashi}\ \emph {et~al.}(2016)\citenamefont
  {Kobayashi}, \citenamefont {Okumura}, \citenamefont {Yamada}, \citenamefont
  {Machida},\ and\ \citenamefont {Aoki}}]{Kobayashi2016_flatband}%
  \BibitemOpen
  \bibfield  {author} {\bibinfo {author} {\bibfnamefont {K.}~\bibnamefont
  {Kobayashi}}, \bibinfo {author} {\bibfnamefont {M.}~\bibnamefont {Okumura}},
  \bibinfo {author} {\bibfnamefont {S.}~\bibnamefont {Yamada}}, \bibinfo
  {author} {\bibfnamefont {M.}~\bibnamefont {Machida}}, \ and\ \bibinfo
  {author} {\bibfnamefont {H.}~\bibnamefont {Aoki}},\ }\href {\doibase
  10.1103/PhysRevB.94.214501} {\bibfield  {journal} {\bibinfo  {journal} {Phys.
  Rev. B}\ }\textbf {\bibinfo {volume} {94}},\ \bibinfo {pages} {214501}
  (\bibinfo {year} {2016})}\BibitemShut {NoStop}%
\bibitem [{\citenamefont {Arrigoni}(1996)}]{Arrigoni1996_3-leg}%
  \BibitemOpen
  \bibfield  {author} {\bibinfo {author} {\bibfnamefont {E.}~\bibnamefont
  {Arrigoni}},\ }\href {\doibase
  http://dx.doi.org/10.1016/0375-9601(96)00215-0} {\bibfield  {journal}
  {\bibinfo  {journal} {Physics Letters A}\ }\textbf {\bibinfo {volume}
  {215}},\ \bibinfo {pages} {91 } (\bibinfo {year} {1996})}\BibitemShut
  {NoStop}%
\bibitem [{\citenamefont {Schulz}(1996)}]{Schulz1996_ladder}%
  \BibitemOpen
  \bibfield  {author} {\bibinfo {author} {\bibfnamefont {H.~J.}\ \bibnamefont
  {Schulz}},\ }\href {\doibase 10.1103/PhysRevB.53.R2959} {\bibfield  {journal}
  {\bibinfo  {journal} {Phys. Rev. B}\ }\textbf {\bibinfo {volume} {53}},\
  \bibinfo {pages} {R2959} (\bibinfo {year} {1996})}\BibitemShut {NoStop}%
\bibitem [{\citenamefont {Kimura}\ \emph {et~al.}(1996)\citenamefont {Kimura},
  \citenamefont {Kuroki},\ and\ \citenamefont {Aoki}}]{Kimura1996_3leg-ladder}%
  \BibitemOpen
  \bibfield  {author} {\bibinfo {author} {\bibfnamefont {T.}~\bibnamefont
  {Kimura}}, \bibinfo {author} {\bibfnamefont {K.}~\bibnamefont {Kuroki}}, \
  and\ \bibinfo {author} {\bibfnamefont {H.}~\bibnamefont {Aoki}},\ }\href
  {\doibase 10.1103/PhysRevB.54.R9608} {\bibfield  {journal} {\bibinfo
  {journal} {Phys. Rev. B}\ }\textbf {\bibinfo {volume} {54}},\ \bibinfo
  {pages} {R9608} (\bibinfo {year} {1996})}\BibitemShut {NoStop}%
\bibitem [{\citenamefont {Lin}\ \emph {et~al.}(1997)\citenamefont {Lin},
  \citenamefont {Balents},\ and\ \citenamefont
  {Fisher}}]{Lin-Balents-Fisher1997_n-leg}%
  \BibitemOpen
  \bibfield  {author} {\bibinfo {author} {\bibfnamefont {H.-H.}\ \bibnamefont
  {Lin}}, \bibinfo {author} {\bibfnamefont {L.}~\bibnamefont {Balents}}, \ and\
  \bibinfo {author} {\bibfnamefont {M.~P.~A.}\ \bibnamefont {Fisher}},\ }\href
  {\doibase 10.1103/PhysRevB.56.6569} {\bibfield  {journal} {\bibinfo
  {journal} {Phys. Rev. B}\ }\textbf {\bibinfo {volume} {56}},\ \bibinfo
  {pages} {6569} (\bibinfo {year} {1997})}\BibitemShut {NoStop}%
\bibitem [{\citenamefont {Lieb}(1989)}]{Lieb}%
  \BibitemOpen
  \bibfield  {author} {\bibinfo {author} {\bibfnamefont {E.}~\bibnamefont
  {Lieb}},\ }\href@noop {} {\bibfield  {journal} {\bibinfo  {journal} {Phys.
  Rev. Lett.}\ }\textbf {\bibinfo {volume} {62}},\ \bibinfo {pages} {1201}
  (\bibinfo {year} {1989})}\BibitemShut {NoStop}%
\bibitem [{\citenamefont {Kontani}\ and\ \citenamefont
  {Ueda}(1998)}]{Kontani_trellis}%
  \BibitemOpen
  \bibfield  {author} {\bibinfo {author} {\bibfnamefont {H.}~\bibnamefont
  {Kontani}}\ and\ \bibinfo {author} {\bibfnamefont {K.}~\bibnamefont {Ueda}},\
  }\href@noop {} {\bibfield  {journal} {\bibinfo  {journal} {Phys. Rev. Lett.}\
  }\textbf {\bibinfo {volume} {80}},\ \bibinfo {pages} {5619} (\bibinfo {year}
  {1998})}\BibitemShut {NoStop}%
\bibitem [{\citenamefont {M\"uller}\ \emph {et~al.}(1998)\citenamefont
  {M\"uller}, \citenamefont {Anisimov}, \citenamefont {Rice}, \citenamefont
  {Dasgupta},\ and\ \citenamefont {Saha-Dasgupta}}]{Muller_SrCu2O3}%
  \BibitemOpen
  \bibfield  {author} {\bibinfo {author} {\bibfnamefont {T.~F.~A.}\
  \bibnamefont {M\"uller}}, \bibinfo {author} {\bibfnamefont {V.}~\bibnamefont
  {Anisimov}}, \bibinfo {author} {\bibfnamefont {T.~M.}\ \bibnamefont {Rice}},
  \bibinfo {author} {\bibfnamefont {I.}~\bibnamefont {Dasgupta}}, \ and\
  \bibinfo {author} {\bibfnamefont {T.}~\bibnamefont {Saha-Dasgupta}},\ }\href
  {\doibase 10.1103/PhysRevB.57.R12655} {\bibfield  {journal} {\bibinfo
  {journal} {Phys. Rev. B}\ }\textbf {\bibinfo {volume} {57}},\ \bibinfo
  {pages} {R12655} (\bibinfo {year} {1998})},\ \bibinfo {note} {in which band
  calculation for SrCu$_2$O$_3$ has been performed}\BibitemShut {NoStop}%
\bibitem [{\citenamefont {Vaugier}\ \emph {et~al.}(2012)\citenamefont
  {Vaugier}, \citenamefont {Jiang},\ and\ \citenamefont
  {Biermann}}]{Vaugier_2012_cRPA_3d-4d}%
  \BibitemOpen
  \bibfield  {author} {\bibinfo {author} {\bibfnamefont {L.}~\bibnamefont
  {Vaugier}}, \bibinfo {author} {\bibfnamefont {H.}~\bibnamefont {Jiang}}, \
  and\ \bibinfo {author} {\bibfnamefont {S.}~\bibnamefont {Biermann}},\ }\href
  {\doibase 10.1103/PhysRevB.86.165105} {\bibfield  {journal} {\bibinfo
  {journal} {Phys. Rev. B}\ }\textbf {\bibinfo {volume} {86}},\ \bibinfo
  {pages} {165105} (\bibinfo {year} {2012})}\BibitemShut {NoStop}%
\bibitem [{\citenamefont {Mravlje}\ \emph {et~al.}(2011)\citenamefont
  {Mravlje}, \citenamefont {Aichhorn}, \citenamefont {Miyake}, \citenamefont
  {Haule}, \citenamefont {Kotliar},\ and\ \citenamefont
  {Georges}}]{Mravlje_2011_cRPA}%
  \BibitemOpen
  \bibfield  {author} {\bibinfo {author} {\bibfnamefont {J.}~\bibnamefont
  {Mravlje}}, \bibinfo {author} {\bibfnamefont {M.}~\bibnamefont {Aichhorn}},
  \bibinfo {author} {\bibfnamefont {T.}~\bibnamefont {Miyake}}, \bibinfo
  {author} {\bibfnamefont {K.}~\bibnamefont {Haule}}, \bibinfo {author}
  {\bibfnamefont {G.}~\bibnamefont {Kotliar}}, \ and\ \bibinfo {author}
  {\bibfnamefont {A.}~\bibnamefont {Georges}},\ }\href {\doibase
  10.1103/PhysRevLett.106.096401} {\bibfield  {journal} {\bibinfo  {journal}
  {Phys. Rev. Lett.}\ }\textbf {\bibinfo {volume} {106}},\ \bibinfo {pages}
  {096401} (\bibinfo {year} {2011})}\BibitemShut {NoStop}%
\bibitem [{\citenamefont {Jang}\ \emph {et~al.}(2016)\citenamefont {Jang},
  \citenamefont {Sakakibara}, \citenamefont {Kino}, \citenamefont {Kotani},
  \citenamefont {Kuroki},\ and\ \citenamefont {Han}}]{Jang_2016_cRPA}%
  \BibitemOpen
  \bibfield  {author} {\bibinfo {author} {\bibfnamefont {S.~W.}\ \bibnamefont
  {Jang}}, \bibinfo {author} {\bibfnamefont {H.}~\bibnamefont {Sakakibara}},
  \bibinfo {author} {\bibfnamefont {H.}~\bibnamefont {Kino}}, \bibinfo {author}
  {\bibfnamefont {T.}~\bibnamefont {Kotani}}, \bibinfo {author} {\bibfnamefont
  {K.}~\bibnamefont {Kuroki}}, \ and\ \bibinfo {author} {\bibfnamefont {M.~J.}\
  \bibnamefont {Han}},\ }\href@noop {} {\bibfield  {journal} {\bibinfo
  {journal} {Scientific reports}\ }\textbf {\bibinfo {volume} {6}},\ \bibinfo
  {pages} {33397} (\bibinfo {year} {2016})}\BibitemShut {NoStop}%
\bibitem [{\citenamefont {Bickers}\ and\ \citenamefont
  {White}(1991)}]{Bickers1991}%
  \BibitemOpen
  \bibfield  {author} {\bibinfo {author} {\bibfnamefont {N.~E.}\ \bibnamefont
  {Bickers}}\ and\ \bibinfo {author} {\bibfnamefont {S.~R.}\ \bibnamefont
  {White}},\ }\href {\doibase 10.1103/PhysRevB.43.8044} {\bibfield  {journal}
  {\bibinfo  {journal} {Phys. Rev. B}\ }\textbf {\bibinfo {volume} {43}},\
  \bibinfo {pages} {8044} (\bibinfo {year} {1991})}\BibitemShut {NoStop}%
\bibitem [{\citenamefont {Sakakibara}\ \emph {et~al.}(2010)\citenamefont
  {Sakakibara}, \citenamefont {Usui}, \citenamefont {Kuroki}, \citenamefont
  {Arita},\ and\ \citenamefont {Aoki}}]{Sakakibara2010}%
  \BibitemOpen
  \bibfield  {author} {\bibinfo {author} {\bibfnamefont {H.}~\bibnamefont
  {Sakakibara}}, \bibinfo {author} {\bibfnamefont {H.}~\bibnamefont {Usui}},
  \bibinfo {author} {\bibfnamefont {K.}~\bibnamefont {Kuroki}}, \bibinfo
  {author} {\bibfnamefont {R.}~\bibnamefont {Arita}}, \ and\ \bibinfo {author}
  {\bibfnamefont {H.}~\bibnamefont {Aoki}},\ }\href {\doibase
  10.1103/PhysRevLett.105.057003} {\bibfield  {journal} {\bibinfo  {journal}
  {Phys. Rev. Lett.}\ }\textbf {\bibinfo {volume} {105}},\ \bibinfo {pages}
  {057003} (\bibinfo {year} {2010})}\BibitemShut {NoStop}%
\bibitem [{3le()}]{3leg_flat_comment}%
  \BibitemOpen
  \href@noop {} {}\bibinfo {note} {The difference between the two-leg and
  three-leg ladders lies in that the two sites within a unit cell is equivalent
  in the former, so that they always have the same weight in the bonding and
  antibonding bands, regardless of the electron-electron interaction. Since the
  flatness of the bands is due to the interference between the two sites, it is
  kept flat even when the interaction is turned on.}\BibitemShut {Stop}%
\bibitem [{dia()}]{diamond_half_comment}%
  \BibitemOpen
  \href@noop {} {}\bibinfo {note} {A similar situation occurs in the diamond
  chain when $t'=1$ and $n\simeq 3$, where the chemical potential lies in the
  vicinity of the bottom of the uppermost band. There, the calculation did not
  converge and $\lambda$ could not be obtained. In fact, this parameter regime
  is beyond the applicability of the FLEX approximation because a Mott
  transition is expected to take place at $n=3$. In any case, we expect absence
  of superconductivity due to the Mottness.}\BibitemShut {Stop}%
\bibitem [{\citenamefont {Misumi}\ and\ \citenamefont
  {Aoki}(2017)}]{Misumi_flatband}%
  \BibitemOpen
  \bibfield  {author} {\bibinfo {author} {\bibfnamefont {T.}~\bibnamefont
  {Misumi}}\ and\ \bibinfo {author} {\bibfnamefont {H.}~\bibnamefont {Aoki}},\
  }\href@noop {} {\bibfield  {journal} {\bibinfo  {journal} {Phys. Rev. B}\
  }\textbf {\bibinfo {volume} {96}},\ \bibinfo {pages} {155137} (\bibinfo
  {year} {2017})}\BibitemShut {NoStop}%
\bibitem [{\citenamefont {Ogura}\ \emph {et~al.}()\citenamefont {Ogura},
  \citenamefont {Aoki},\ and\ \citenamefont {Kuroki}}]{Ogura_Hiddenladder}%
  \BibitemOpen
  \bibfield  {author} {\bibinfo {author} {\bibfnamefont {D.}~\bibnamefont
  {Ogura}}, \bibinfo {author} {\bibfnamefont {H.}~\bibnamefont {Aoki}}, \ and\
  \bibinfo {author} {\bibfnamefont {K.}~\bibnamefont {Kuroki}},\ }\href@noop {}
  {\bibfield  {journal} {\bibinfo  {journal} {arXiv preprint arXiv:1706.01711}\
  }}\bibinfo {note} {, to be published in Phys. Rev. B}\BibitemShut {NoStop}%
\bibitem [{\citenamefont {Nakata}\ \emph {et~al.}(2017)\citenamefont {Nakata},
  \citenamefont {Ogura}, \citenamefont {Usui},\ and\ \citenamefont
  {Kuroki}}]{Nakata2017_finite-energy}%
  \BibitemOpen
  \bibfield  {author} {\bibinfo {author} {\bibfnamefont {M.}~\bibnamefont
  {Nakata}}, \bibinfo {author} {\bibfnamefont {D.}~\bibnamefont {Ogura}},
  \bibinfo {author} {\bibfnamefont {H.}~\bibnamefont {Usui}}, \ and\ \bibinfo
  {author} {\bibfnamefont {K.}~\bibnamefont {Kuroki}},\ }\href {\doibase
  10.1103/PhysRevB.95.214509} {\bibfield  {journal} {\bibinfo  {journal} {Phys.
  Rev. B}\ }\textbf {\bibinfo {volume} {95}},\ \bibinfo {pages} {214509}
  (\bibinfo {year} {2017})}\BibitemShut {NoStop}%
\end{thebibliography}%

\end{document}